\documentclass[twocolumn,aps,prx,reprint,twocolumn,amsmath,amssymb]{revtex4-1}
\usepackage[bookmarks=false,linkcolor=blue,urlcolor=blue,colorlinks,citecolor=blue]{hyperref}
\usepackage{amsmath}
\usepackage{graphicx}
\begin{document}
\title{Topological superconductivity in carbon nanotubes with a small magnetic flux}
\author{Omri Lesser}
\thanks{These authors contributed equally to the work.}
\affiliation{Department of Condensed Matter Physics, Weizmann Institute of Science,
Rehovot, Israel 76100}
\author{Gal Shavit}
\thanks{These authors contributed equally to the work.}
\affiliation{Department of Condensed Matter Physics, Weizmann Institute of Science,
Rehovot, Israel 76100}
\author{Yuval Oreg}
\affiliation{Department of Condensed Matter Physics, Weizmann Institute of Science,
Rehovot, Israel 76100}
\begin{abstract}
We show that a one-dimensional topological superconductor can be realized
in carbon nanotubes, using a relatively small magnetic field. Our
analysis relies on the intrinsic curvature-enhanced spin-orbit coupling
of the nanotubes, as well as on the orbital effect of a magnetic flux
threaded through the nanotube. Tuning experimental parameters, we show that
a half-metallic state may be induced in the nanotube. Coupling the
system to an Ising superconductor, with an appreciable spin-triplet
component, can then drive the nanotube into a topological superconducting
phase. The proposed scheme is investigated by means of real-space
tight-binding simulations, accompanied by an effective continuum low-energy
theory, which allows us to gain some insight on the roles of different
terms in the Hamiltonian. We calculate the topological phase diagram
and ascertain the existence of localized Majorana zero modes near
the edges. Moreover, we find that in the absence of
a magnetic field, a regime exists where sufficiently strong interactions
drive the system into a time-reversal-invariant topological superconducting
phase.
\end{abstract}
\maketitle
\section{Introduction}

Low-dimensional topological superconductors are unique states of matter,
supporting Majorana fermions at the system's edges~\cite{ReadGreen,Kitaev_2001,LutchynMajorana,OregMajoran}.
These zero-energy edge modes have non-Abelian exchange statistics,
making them a very attractive platform for realizing quantum computation
schemes~\cite{KITAEV20032,NayakStern}. Experimental evidence for
the emergence of Majorana zero modes, at the ends of one-dimensional
(1D) semiconducting nanowires with strong Rashba spin-orbit coupling (SOC) and induced
Zeeman spin splitting, was observed in the form of zero-bias conductance
peaks in several instances~\cite{Deng2012,Mourik1003,Das2012,MarcusMajoranas,Lutchyn2018Review}.

An alternative route to realizing 1D topological superconductivity
is using carbon nanotubes~\cite{Iijima1991} (CNTs) instead of semiconducting
nanowires. CNTs are small-diameter tubes of rolled-up graphene, having
exceptional electronic band structures and transport properties~\cite{RMPcnt,dresselhaus_physics_1995}.
As opposed to nanowires, CNTs have a truly-1D nature, as their diameter
$d$ is extremely small (of order $1\,\text{nm}$). Moreover, being comprised entirely of carbon atoms, very clean CNTs
may be fabricated, thus facilitating probing of their quantum properties~\cite{RMPcntcleantransport}.

These properties make CNTs an attractive platform for pursuing 1D
Majorana fermions, and several schemes aimed at achieving those have been put forward~\cite{SauMajoranCNT,GrifoniMajoranaCNT,CNTmajoranaLoss}.
The proposals mainly rely on the same ingredients available in the
semiconducting-nanowires setups: a combination of proximity to an
$s$-wave superconductor, SOC, and a Zeeman magnetic field. The latter,
due to the low $g$-factor of the CNTs, typically needs to be very
large, which poses experimental challenges: high magnetic fields are
not easily produced, and may also critically suppress superconductivity
in the proximitizing substrate.

\begin{figure}
\begin{centering}
\includegraphics[scale=0.49]{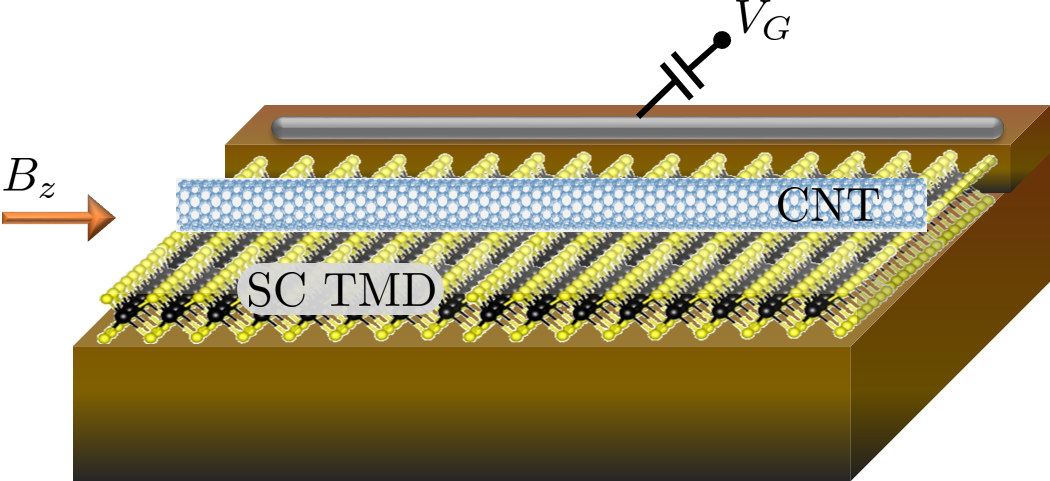}
\par\end{centering}
\caption{\label{fig:1schematic} Our proposed setup for realizing topological
superconductivity in a CNT. We apply a magnetic field $B_{z}$ parallel
to the CNT, and use a nearby metallic gate (gray) with voltage $V_{\rm G}$
to tune its chemical potential. In the presence of SOC and rotational-symmetry breaking (inherent in our setup), these allow us to tune
the CNT to a half-metallic point. Thus, proximity coupling to a superconducting
TMD substarte may open a topological gap in the CNT, hosting Majorana
states at the edge.}
\end{figure}

In this manuscript, we present a scheme which allows us to circumvent
the high Zeeman-energy problem, and realize topological Majorana zero
modes \textit{without the need for any Zeeman splitting}. Our scheme,
depicted in Fig.~\ref{fig:1schematic}, relies instead on an orbital
effect caused by a magnetic flux threaded through the nanotube. When it is combined with the unusual SOC present in CNTs, and
the breaking of the CNT's rotational symmetry (by, e.g., an external
gate), the CNT can be tuned into a half-metallic state using relatively
low magnetic fields. Then, proximitizing the CNT to a superconductor
with a significant spin-triplet component in its Cooper-pairs wavefunction,
a $p$-wave topological gap may open in the nanotube, hosting Majorana
fermions near its edges. Thin films of transition-metal dichalcogenides
(TMDs) make excellent candidates for the superconducting substrate,
having strong Ising SOC, favorable for pairing of electrons with their spin polarized
in the TMD plane.

The presence of spin-triplet pairs in the superconducting
substrate opens up another interesting possibility, as the interaction
between electrons in the CNT heavily favors triplet  pairing over
the singlets.
Then, in a regime with zero magnetic flux, strong enough interactions may allow one to tune the system into a time-reversal-invariant topological
superconductor phase~\cite{TRITOPSinvariant,TritopsReview}.

The rest of the manuscript is organized as follows. In Sec.~\ref{Model},
we introduce the theoretical model of our system, as well as the effective low-energy theory.
Sec.~\ref{Proximity} is dedicated to the spin-triplet proximity
effect, and its implementation using superconducting TMDs. We show
that a topological superconducting phase is supported by our model
in Sec.~\ref{Results}. The presence of Majorana zero modes bound to the edges is demonstrated in Sec.~\ref{Majoranas}.
The conditions for realizing a topological phase without magnetic flux are presented and discussed in Sec.~\ref{triptopssec}.
We conclude our findings in Sec.~\ref{Conclus}.

\section {Model for the CNT}\label{Model}

We consider a tight-binding model of the $\pi$-electrons of a cylindrically
rolled-up graphene lattice comprising the CNT. The CNT may be classified
by its chiral vector $\mathbf{C}=\left(n,m\right)$, describing the
rolling direction in the hexagonal-lattice plane. The resulting spectrum
then includes a series of 1D ``cuts'' of the 2D Dirac cones, which
are determined by the chiral vector~\cite{dresselhaus1998physical}.
We focus in this work on metallic zigzag nanotubes, i.e., CNTs where
$\mathbf{C}=\left(n,0\right)$ and $n\in3\mathbb{Z}$, yet our model
is easily generalized to any metallic zigzag-like CNT with $n,m\in3\mathbb{Z}$
and $n\neq m$, as we discuss below.
Importantly, for this kind of CNTs the pure hopping spectrum (without,
e.g., SOC) is gapless and four-fold degenerate ($2$ spin $\times$
$2$ valley) near $k_{\parallel}=0$, where $k_{\parallel}$ is the
momentum along the nanotube axis.

The CNT is modeled by the following tight-binding Hamiltonian on a
honeycomb lattice with periodic boundary conditions in a direction
determined by $\mathbf{C}$,
\begin{equation}
\begin{aligned}H_{\text{CNT}} & =\sum_{i,s,s'}c_{i,s}^{\dagger}\left(-\delta^{ss'}\mu\left(\theta_{i}\right)-\sigma_{z}^{ss'}V_{\rm Z}\right)c_{i,s'}\\
+ & \sum_{\left\langle i,j\right\rangle ,s,s'}\left[c_{i,s}^{\dagger}\left(-t\delta^{ss'}e^{iA_{ij}}+i\Delta_{o,ij}^{\text{SO}}\sigma_{z}^{ss'}\right)c_{j,s'}+{\rm h.c.}\right]\\
+ & \sum_{\left\langle \left\langle i,j\right\rangle \right\rangle ,s,s'}\left[i\Delta_{z,ij}^{\text{SO}}c_{i,s}^{\dagger}\left(\sigma_{z}\right)_{ss'}c_{j,s'}+{\rm h.c.}\right].
\end{aligned}
\label{eq:H_CNT_tight_binding}
\end{equation}
Here $c_{i,s}$ are creation operators of electrons at the lattice
site $i$ with spin $s$, $t$ is the nearest-neighbor hopping amplitude,
$\mu$ is the on-site chemical potential, which in general depends
on the angle $\theta_{i}$ along the CNT's circumference at which
the site $i$ is situated, and $\sigma_{z}$ is a Pauli matrix acting
in spin space. The magnetic field applied along the CNT axis gives
rise to the Zeeman splitting $V_{\rm Z}$ and to an orbital effect, captured
by the Peierls phase $A_{ij}$~\cite{peierls_zur_1933}. The SOC is accounted for by two terms,
$\Delta_{o,ij}^{\text{SO}},\Delta_{z,ij}^{\text{SO}}$, which are
the orbital- and Zeeman-type SOC matrix elements between sites $i,j$.
The labels $\left\langle i,j\right\rangle $ and $\left\langle \left\langle i,j\right\rangle \right\rangle $
indicate summation over nearest and next-nearest neighbors, respectively.
The on-site potential $\mu\left(\theta_{i}\right)$ breaks the azimuthal
symmetry of the CNT, enabling inter-valley scattering, which will
prove crucial for our subsequent analysis. More details regarding
the tight-binding model Eq.~\eqref{eq:H_CNT_tight_binding} are given in Appendix~\ref{appendix:TB_details}.

\begin{figure*}
\begin{centering}
\includegraphics[width=1\textwidth]{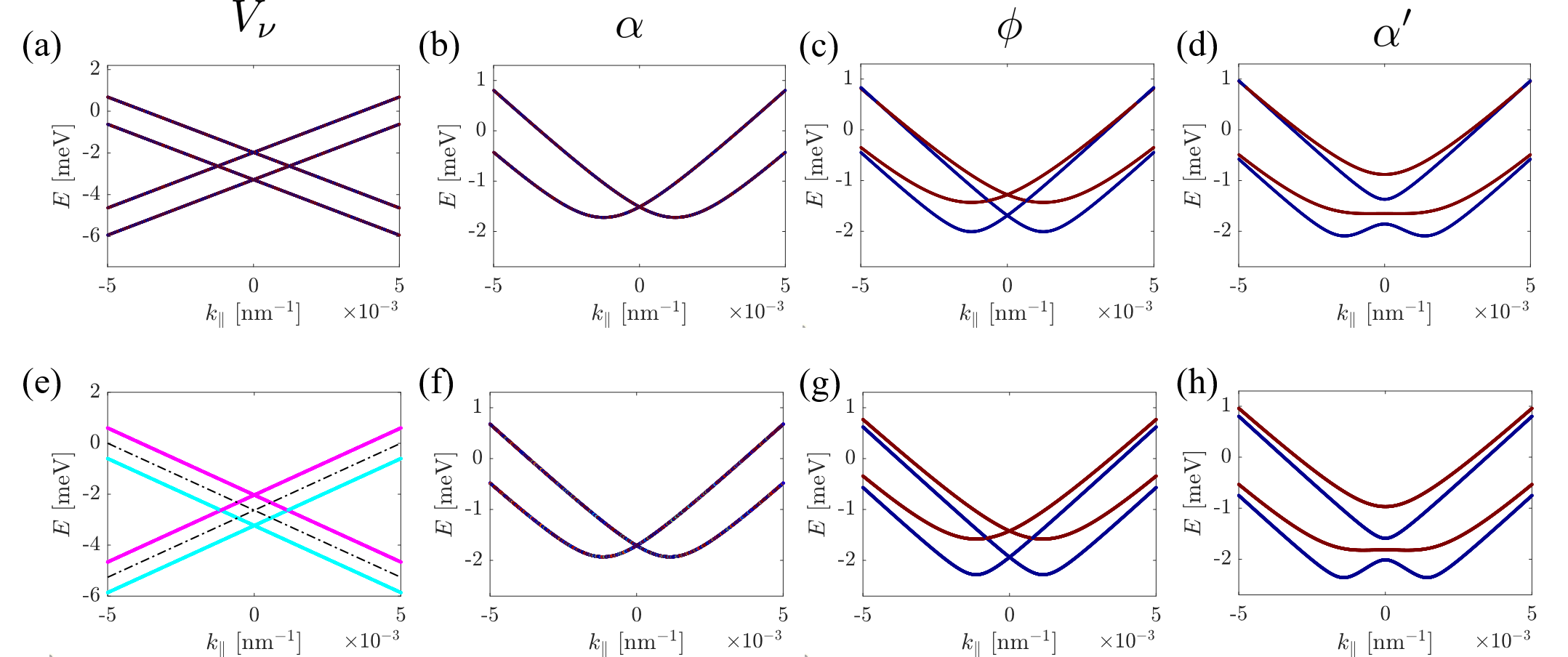}
\par\end{centering}
\caption{\label{fig:2SpectraUnited} Spectra of a $\left(12,0\right)$ zigzag CNT calculated using the tight-binding Hamiltonian Eq.~\eqref{eq:H_CNT_tight_binding}
(top row), and using the effecive continuum model Eq.~\eqref{eq:continuumNormalH}
(bottom row). Going from left to right, we consecutively ``switch
on'' different terms in the Hamiltonian. (a) Calculated spectrum with
zero magnetic flux and zero SOC terms. The Dirac cone splitting is
due to a rotational symmetry breaking by the on-site chemical potential,
$\mu\left(\theta_{i}\right)=\frac{\mu_{0}}{\sqrt{2\pi}\Sigma}e^{-\frac{1}{2}\left(\frac{\theta_{i}}{\Sigma}\right)^{2}}$,
with $\mu_{0}=40$ meV, and $\Sigma=0.6$. (b) Same as (a), but with
a finite orbital SOC energy, $\Delta_{o}^{\text{SO}}=0.7$ meV. (c)
Same as (b), with an added magnetic flux induced by a magnetic field $B_{z}=1$ T applied along the CNT axis. (d) Same as (c), with a finite Zeeman-type SOC,
$\Delta_{Z}^{\text{SO}}=0.5$ meV. The plots (e-h) correspond to the
effective continuum Hamiltonian of plots (a-d), respectively. The
roles of rotational-symmetry breaking, orbital SOC, Zeeman SOC, and
magnetic flux are captured by the parameters $V_{\nu}=0.6$ meV, $v_{\rm F}\alpha=0.7$
meV, $v_{\rm F}\alpha'=0.5$ meV, and $v_{\rm F}\phi=0.35$ meV, respectively.
In (e) the different colors mark the different expectation values
of the valley bonding and anti-bonding $\left\langle \nu_{x}\right\rangle =\pm1$
in the two bands. In all other plots red and blue mark the spin projections
of the different bands along the nanotube axis, $\sigma=\pm1$. The
tight-binding model and the continuum model agree well, and as expected
deviations start to appear when moving away from $k_{\parallel}=0$.
}
\end{figure*}

We find that the low-energy properties of the tight-binding model
Eq.~\eqref{eq:H_CNT_tight_binding} can be approximated by the following
continuum model (we set $\hbar=1$ henceforth):
\begin{equation}
H=v_{\rm F}\left[\rho_{y}k_{\parallel}+\rho_{x}\nu_{z}\left(\alpha\sigma_{z}+\phi\right)+\alpha'\sigma_{z}\nu_{z}\right]+V_{\nu}\nu_{x},\label{eq:continuumNormalH}
\end{equation}
with $\rho_{i}$, $\sigma_{i}$, $\nu_{i}$ Pauli matrices acting
in the subspaces of the sub-lattice, spin, and valley degrees of freedom, respectively.
The Fermi velocity $v_{\rm F}$ characterizes the linear dispersion near the graphene Dirac cones, with a value of $\sim8\cdot10^{5}$
m/sec. The spin-orbit term $\alpha$ corresponds to a spin-dependent
phase accumulated by an electron going around the tube's diameter~\cite{AndoSOC}.
Its strength is inversely proportional to the diameter of the nanotube~\cite{BrataasSOC}
and can be roughly estimated~\cite{IlaniSOC} as
$v_{\rm F}\alpha\approx1\frac{\rm meV}{R\left[{\rm nm}\right]},$with
$R$ the CNT's radius. The strength of the sub-lattice diagonal ``Zeeman-like''
SOC term $\alpha'$ depends on the chirality of the CNT~\cite{alphaTag},
and is usually estimated to be of the same order of magnitude as $\alpha$~\cite{alphaTag,alphaTag2,SauMajoranCNT}.
The circumferential momentum shift due to the Aharonov-Bohm (AB) flux
$\phi$ can be written in terms of $R$ and the magnetic field $B_{z}$,
$v_{\rm F}\phi\approx R\left[{\rm nm}\right]B_{z}\left[{\rm T}\right]$ meV.

Finally, $V_{\nu}$, which is responsible for inter-valley scattering
due to breaking of the CNT rotational symmetry about its axis, caused
by, e.g., an anisotropic gate inducing angle-dependent chemical potential,
is approximated from our tight-binding analysis to be of order $\sim1$
meV, consistent with previous estimates~\cite{GrifoniMajoranaCNT}.
In Eq.~\eqref{eq:continuumNormalH} the Zeeman term induced by the
external field $B_{z}$ was neglected, as it is small in comparison
to the other energy scales for the moderate-to-low magnetic field
regime we are interested in (few Tesla or lower). For example, with
a magnetic field $B_{z}=1\,{\rm T}$ we have $V_{Z}\approx0.1\,{\rm meV}$,
whereas the energy associated with the flux for $R=1\,{\rm nm}$ is
$V_{\phi}=v_{\rm F}\phi\approx0.85\,{\rm meV}$.
For further details on numerically estimating the parameters appearing in Eq.~\eqref{eq:continuumNormalH} in terms of experimental parameters, see Appendix~\ref{app:cheatsheet}.

It is instructive to define an anti-unitary time-reversal operator
$\mathcal{T}=\nu_{x}\sigma_{y}\mathcal{K}$, with $\mathcal{K}$ the
complex conjugation operator, such that $\mathcal{T}^{2}=-1$. Neglecting
the Zeeman term, only the AB flux term breaks the time-reversal symmetry,
since $H\left(\phi=0\right)$ commutes with $\mathcal{T}$, as one would
expect. Also notice that in the Hamiltonian Eq.~\eqref{eq:continuumNormalH}
the spin projection along the CNT axis, $\sigma_{z}$, is a good quantum
number, to be labeled as $\sigma=\pm1$.

The role of each of the components of $H$ is illustrated in Fig.~\ref{fig:2SpectraUnited}.
The low-energy continuum Hamiltonian is readily diagonalized, and
we obtain the eigen-energies $E=\pm\epsilon\left(k_{\parallel}\right)$,
with\begin{widetext}
\begin{equation}
\epsilon\left(k_{\parallel}\right)=\sqrt{k_{\parallel}^{2}+V_{\nu}^{2}+\left(\phi+\sigma\alpha\right)^{2}+\alpha'^{2}\pm2\sqrt{\left(k_{\parallel}V_{\nu}\right)^{2}+\alpha'^{2}\left[k_{\parallel}^{2}+\left(\phi+\sigma\alpha\right)^{2}\right]}}.\label{eq:spectrumEps}
\end{equation}
\end{widetext}
Several key insights may be inferred from the form
of $\epsilon\left(k_{\parallel}\right)$. First, it is evident that
only a combination of the magnetic flux $\phi$ and the SOC $\alpha$
terms may lift the spin degeneracy in the spectrum, which is vital
for our half-metallic construction. We also see that $V_{\nu}$ splits
the spectrum into two shifted copies in the $k_{\parallel}$ direction,
similar to the effect of Rashba SOC in quantum nanowires. Finally,
the role of Zeeman-like SOC $\alpha'$ is clearly understood in the
vicinity of $k_{\parallel}=0$, where it lifts the two-fold degeneracy
in the spectrum, thereby opening a gap. The lifting of this degeneracy
is also crucial, otherwise one always ends up with an even number
of pairs of Fermi points, regardless of the value of the chemical potential. To
achieve an odd number of Fermi points, and hence the possibility
of a topological phase, one must thus use a CNT which has a finite $\alpha'$ SOC term. 

\begin{figure}
\begin{centering}
\includegraphics[scale=0.66]{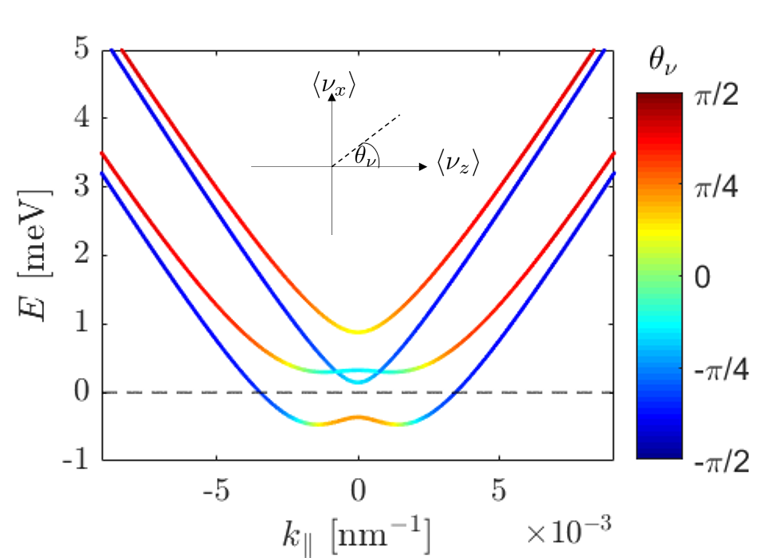}
\par\end{centering}
\caption{\label{fig:3nuz} Zoom-in on part of the CNT spectrum, with the chemical
potential tuned to a half-metallic point. The black dashed line marks
the Fermi energy, and colors indicate the value of $\theta_{\nu}\equiv\tan^{-1}\frac{\left\langle \nu_{x}\right\rangle }{\left\langle \nu_{z}\right\rangle }$
(illustrated in the inset); recall that the $\nu_{i}$ act in the
valley subspace, so $\theta_{\nu}=0$ corresponds to complete valley
polarization, which occurs when the rotation-symmetry breaking term
$V_{\nu}$ vanishes and $\nu_{z}$ is a good quantum number. Here,
$V_{\nu}$ is sufficiently large such that the valleys are almost
completely mixed near the Fermi points, i.e., $\left|\theta_{\nu}\right|\approx\frac{\pi}{2}$.
Parameters used: $v_{\rm F}\alpha=0.4$ meV, $v_{\rm F}\alpha'=0.3$ meV,
$v_{\rm F}\phi=2$ meV, and $V_{\nu}=1$ meV.}
\end{figure}

A single-channel half-metallic phase is achieved when the spectrum
$\epsilon\left(k_{\parallel}\right)$ is tuned such that an energy
window with only two Fermi points exist.  However, this is not sufficient
to ensure that the CNT is susceptible to proximity-induced superconductivity.
The Cooper pair that tunnels from the superconductor typically has
a small net momentum, and therefore the sum of the two Fermi momenta
should also be small. This cannot be achieved if the two Fermi points
belong to the same valley, in which case the total momentum of the pair
in the circumferential direction $k_{\perp}\sim1/R$ is large. This
problem is circumvented in our scheme by introducing a potential that breaks
the symmetry around the tube. This symmetry breaking is embodied by
the term $V_{\nu}\nu_{x}$ in Eq.~\eqref{eq:continuumNormalH}. When
the value of this term is comparable to the other terms in the Hamiltonian,
an appreciable valley mixing is obtained so that $\left|\left\langle \nu_{x}\right\rangle \right|\approx1$
and the two Fermi points have opposite momenta, see Fig.~\ref{fig:3nuz}.
Tuning the chemical potential, such that an additional even number of spin channels is occupied, may also lead to topological
superconductivity.

\section {Equal-spin proximity effect}\label{Proximity}

So far, we have established the possibility of tuning the CNT into
a state where it has an odd number of pairs of Fermi crossing points,
by exploiting the intrinsic SOC and the orbital effect of a parallel
magnetic field. However, an $s$-wave superconductor proximity coupled
to the nanotube cannot induce a topological gap, since all the bands
are spin polarized. Instead, one needs to use a superconductor which
has a significant \textit{spin-triplet} component, and bring it to
contact with the CNT. Moreover, this superconductor should have the
\textit{right} spin-triplet component, that will be compatible with
the spin polarization of the CNT, which is in the tube axis direction.

We propose the use of superconducting thin films or monolayers of
TMDs as a superconducting substrate. In these materials, due to a
combination of strong atomic SOC and breaking of the lattice in-plane
mirror symmetry, electrons in opposite valleys experience opposite
effective Zeeman fields~\cite{TMDsoc1,TMDsoc2}, an effect known
as Ising SOC. Studies of superconducting few-layers TMDs show an increase
of the upper critical in-plane magnetic field well above the Clogston
limit (where the magnetic polarization energy is equal to the superconductor
condensation energy)~\cite{Mos2IsingSC,LargeHcMos2,ISINGtHINnBSE2,Nbse2SpectroscopySteinberg,Nbse2IsingSCtuningHunt,NbSe2Mak}.
This phenomenon originates in the strong tendency of the electron
spins to point in the out-of-plane direction due to a strong Ising
effective field. It was demonstrated that in the presence of an $s$-wave
pairing potential, Ising SOC facilitates equal-spin spin-triplet Cooper
pairs, with their spin pointing in the in-plane direction~\cite{TMDhalfmetal}. This scenario
is ideal for inducing topological superconductivity in the CNT. Notice
that no time-reversal-symmetry breaking within the superconducting
TMD needs to occur. Concretely, we suggest the use of one particular
material, $\text{NbSe}_{2}$, which has an exceptionally high Ising
SOC with a spin-splitting energy of about $80$ meV in the monolayer~\cite{ISINGtHINnBSE2}.
This would ensure that the equal-spin component in the Cooper-pair
wavefunction is comparable with that of the singlet. Moreover, recent
experiments with graphene-superconducting $\text{NbSe}_{2}$ heterostructures imply some compatibility between the two, and the possibility of an
appreciable proximity effect~\cite{Nbse2GrapheneJunction2,NbSe2garaphene3,Nbse2GrapheneJunction},
which will presumably also hold true for the CNTs.

We model the proximity-induced pairing terms in the nanotube as
\begin{equation}
\begin{aligned}H_{\text{SC}} & =\tilde{\Delta}_{s}\sum_{i}\gamma_{i}c_{i\uparrow}^{\dagger}c_{i,\downarrow}^{\dagger}\\
 & +\tilde{\Delta}_{t}\sum_{\left\langle i,j\right\rangle ,s=\uparrow,\downarrow}b_{ij}\gamma_{i}\gamma_{j}c_{i,s}^{\dagger}c_{j,s}^{\dagger}+{\rm h.c.},
\end{aligned}
\label{eq:tbSC}
\end{equation}
where $\tilde{\Delta}_{s},\tilde{\Delta}_{t}$ are the singlet and
triplet pairing potentials, the indicator $\gamma_{i}$ is 1 if site $i$ lies in
the area covered by the SC and 0 otherwise, and $b_{ij}=\pm1$ depending
on the direction of the bond connecting sites $i,j$. The pairing
term Eq.~\eqref{eq:tbSC} can also be captured by the low-energy
continuum description. We introduce an anti-unitary particle-hole
operator $\Lambda=\tau_{y}\mathcal{T}$, with $\tau_{i}$ Pauli matrices
acting on the particle-hole degree of freedom (notice that in the
tight-binding description the valley degree of freedom is absent,
and thus the particle-hole operator is $\Lambda_{\text{TB}}=\tau_{x}\mathcal{K}$).
We may now write the continuum Bogoliubov-de Gennes (BdG) Hamiltonian,
\begin{align}
H_{{\rm BdG}} & =\left\{ v_{\rm F}\left[\rho_{y}k_{\parallel}+\left(\rho_{x}\alpha+\alpha'\right)\sigma_{z}\nu_{z}\right]+V_{\nu}\nu_{x}-\mu\right\} \tau_{z}\nonumber \\
 & +v_{\rm F}\rho_{x}\nu_{z}\phi+\left(\Delta_{s}+\Delta_{t}\rho_{y}\sigma_{x}\right)\tau_{x},\label{eq:bdgHamiltonian}
\end{align}
which acts on a Nambu spinor with a total of 16 components (sub-lattice,
spin, valley, and particle-hole). $\Delta_{s}$,$\Delta_{t}$ are
the low-energy counterparts of the tight-binding pairing potentials
introduced in Eq.~\eqref{eq:tbSC}, $\tilde{\Delta}_{s},\tilde{\Delta}_{t}$,
respectively. The particle-hole symmetry is manifested by $\left\{ H_{{\rm BdG}},\Lambda\right\} =0$.
The form of the spin-triplet term in $H_{{\rm BdG}}$ is not only
consistent with the tight-binding simulations, but it is also the \textit{only}
possible pairing term which (i) preserves particle-hole and time-reversal
symmetries, (ii) pairs nearest-neighbor equal-spin electrons with zero
circumferential momentum, and (iii) does not distinguish between different
valleys. Although the applied magnetic field breaks the time-reversal
symmetry in our system, the intrinsic pairing in the superconducting
substrate does not. The same argument applies with regards to the
breaking of the valley symmetry by $V_{\nu}$. Thus, this pairing
term, along with the singlet one, is the main focus in this work.
We note that in the absence of singlet pairing, $\Delta_{s}=0$, the
spin conservation in the system is reflected by $\left[H_{{\rm BdG}},\sigma_{z}\tau_{z}\right]=0$,
i.e., $\sigma_{z}$ is no longer a good quantum number, but
$\sigma_{z}\tau_{z}$ is.\textcolor{red}{{} }A finite $\Delta_{s}$
breaks this symmetry as it mixes the spins, but is not necessarily
detrimental to the emergence of the topological superconducting phase, as we will later show.

A lattice BdG Hamiltonian can be used to diagnose the parameter regimes where topological superconductivity
takes place by introducing the $\mathbb{Z}_{2}$ topological index~\cite{Kitaev_2001,PfaffianInvariant,PfaffianMultiband},
\begin{equation}
Q={\rm sgn}\left[{\rm Pf}\left\{ \Lambda H_{\text{BdG}}\left(k_{\parallel}=0\right)\right\} {\rm Pf}\left\{ \Lambda H_{\text{BdG}}\left(k_{\parallel}=\pi\right)\right\} \right],\label{eq:PfaffaianInvariant}
\end{equation}
where ${\rm Pf}\left\{ \cdot\right\} $ is the matrix Pfaffian. $Q=-1$
corresponds to the topologically non-trivial phase whereas in the
trivial phase $Q=1$. For the case of a continuum BdG Hamiltonian as in Eq.~\eqref{eq:bdgHamiltonian}, the Pfaffian should only be evaluated at $k_{\parallel}=0$.

\section {Emergence of topological superconductivity}\label{Results}

\begin{figure}
\begin{centering}
\includegraphics[scale=0.67]{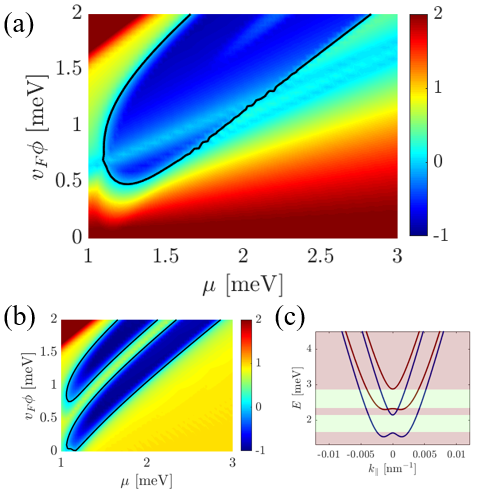}
\par\end{centering}
\caption{\label{fig:4muphis} (a) Phase diagram of the proximity-coupled CNT
as a function of the magnetic flux and chemical potential. Shown is
the BdG quasi-particle energy gap, normalized by the triplet pairing
strength, $\frac{E_{\rm g}}{\Delta_{t}}$, multiplied by the topological
index $Q$. Regions with negative values (blue) are topological. The
black lines demarcate the topological phase transitions, where the
gap closes. The parameters used are $v_{\rm F}\alpha=0.4$ meV, $v_{\rm F}\alpha'=0.3$
meV, $V_{\nu}=1$ meV, $\Delta_{t}=0.1$ meV, and $\Delta_{s}=0.3$
meV. (b) The same as (a), only with the spin-singlet component of
the proximity effect $\Delta_{s}=0$. (c) The normal-state CNT spectrum,
with the same parameters as in Fig.~\ref{fig:3nuz}, with different
line colors for the two spin directions. We mark the region with an
odd (even) number of Fermi crossing pairs with a bright green (pink)
background. The small pink ``window'' in between the bright green
regions corresponds to the trivial phase in between the two topological
ones in (b).}
\end{figure}

Now that we have the full BdG Hamiltonian Eq.~\eqref{eq:bdgHamiltonian},
we may explore the parameter space in order to find the topological
phases. As an example, using realistic parameters for the CNT, as
well as for the superconducting TMD substrate, we show in Fig.~\ref{fig:4muphis}a
that by ``scanning'' the gate voltage and the magnetic flux
a large topological area in parameter space is indeed accessible,
with a quasi-particle gap $E_{\rm g}$ comparable in size to the triplet
pairing potential $\Delta_{t}$.

Interestingly, while $\Delta_{t}$ is the crucial ingredient for our
scheme to produce topological $p$-wave superconductivity, the presence
of a finite singlet component $\Delta_{s}$ can be beneficial in some
cases. This can be seen by comparing Fig.~\ref{fig:4muphis}a and
Fig.~\ref{fig:4muphis}b, where in the latter $\Delta_{s}=0$, and
a trivial regime emerges in the middle of the area which was topological in the former.
To understand why, one should examine the number of Fermi level
crossings in the normal-state spectrum, see Fig.~\ref{fig:4muphis}c.
When varying, e.g., the chemical potential, a crossover occurs from
an odd number of crossing pairs to an even one. Having only same-spin
pairing would thus mean we have a topological phase transition to
the trivial phase. A finite $\Delta_{s}$ may however bridge between
the two topological phases by allowing more pairing interactions in
the intermediate region. Another consequence of finite $\Delta_{s}$
is an increase of the minimal magnetic field required to access the
topological phase. This is to be expected, since $\phi$ is necessary
to establish the half-metallic phase, in which there exists a regime
where the spin-singlet pairing is ineffective.

Let us comment on the strength of the magnetic field required to tune the system into the topological phase.
The SOC parameters chosen for Fig.~\ref{fig:4muphis} are appropriate for a CNT of radius 2.5nm. 
The minimal $v_{\rm F} \phi$ required to make this CNT topological is about 0.5meV, which by the relation $B_z[T] \approx v_{\rm F} \phi[\text{meV}]/R[\text{nm}]$ corresponds to a magnetic field of 200mT (see Appendix~\ref{app:cheatsheet}).
In this example we find that the gap is $E_{\rm g} \approx 0.1$meV. The localization length of the MZM is $\xi \approx v_{\rm F}/E_{\rm g} \approx 5\mu\text{m}$, and we expect that it may be smaller due to a reduction of the Fermi velocity by the superconductor~\cite{peng_strong_2015}. Following this consideration we estimate that a nanotube with a radius of 10nm requires only about 50mT to become topologically nontrivial, albeit with a smaller gap.

\begin{figure}
\begin{centering}
\includegraphics[scale=0.65]{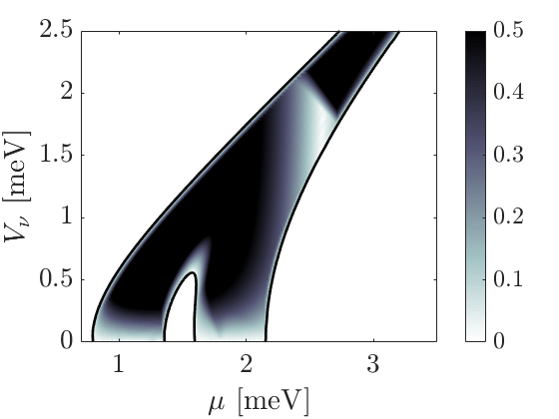}
\par\end{centering}
\caption{\label{fig:5muVnu} Normalized quasi-particle energy gap $\frac{E_{\rm g}}{\Delta_{t}}$,
as a function of the chemical potential $\mu$ and the inter-valley
mixing strength $V_{\nu}$. The color scale is such that only the
variations inside the topological regime are shown. Notice that the gap
vanishes as $V_{\nu}\rightarrow0$. The parameters used are $v_{\rm F}\alpha=0.4$
meV, $v_{\rm F}\alpha'=0.3$ meV, $v_{\rm F}\phi=1.5$ meV, $\Delta_{t}=0.1$
meV, and $\Delta_{s}=0.3$ meV.}
\end{figure}

We can now also better appreciate the role of $V_{\nu}$ in $H_{{\rm BdG}}$.
Examining the amplitude of the topological gap as a function of $V_{\nu}$,
see Fig.~\ref{fig:5muVnu}, we see that near $V_{\nu}=0$ there exists
a region which although being formally topological as $Q=-1$, has
a very small energy gap $E_{\rm g}$ -- it is diminished by around one
order of magnitude compared to $\Delta_{t}$. This is merely a consequence
of the valley polarization in the absence of the rotational-symmetry
breaking, which attenuates zero-momentum Cooper-pair hopping into
the CNT. Only when the average valley number $\left\langle \nu_{z}\right\rangle $
approaches zero, namely the electron wavefunctions at the Fermi surface
are evenly distributed between $K$ and $K'$, can the full superconducting
gap develop in the system, as is the case for larger values of $V_{\nu}$.

\section {Majorana edge states}\label{Majoranas}

We now turn to demonstrate the topological phase transition from real-space diagonalization of the tight-binding Hamiltonian Eqs.~\eqref{eq:H_CNT_tight_binding},~\eqref{eq:tbSC}.
To this end, we simulate a finite-length CNT with open boundary conditions.
We exemplify our results on a $\left(6,0\right)$ CNT of length $12.3\ \mu\text{m}$
($1.2\cdot10^{6}$ carbon atoms). We use the realistic parameters
$t=2.66\text{eV}$~\cite{tomanek_first-principles_1988}, $\Delta_{o}^{\text{SO}}=2\text{meV}$,
$\Delta_{z}^{\text{SO}}=1\text{meV}$~\cite{IlaniSOC}, and assume
an inter-valley mixing energy of $15\text{\,meV}$ in a step-like
structure (see Appendix~\ref{appendix:TB_details} for details).
A modest magnetic field of $2\text{T}$ is used in order to drive
the system into the topological phase (for CNTs of larger diameter,
even a weaker magnetic field will suffice). For simplicity, we discard
the spin-singlet component of the SC $\tilde{\Delta}_{s}=0$, and
only include a spin-triplet component $\tilde{\Delta}_{t}=0.5\text{meV}$.

\begin{figure}
\begin{centering}
\includegraphics[width=1\linewidth]{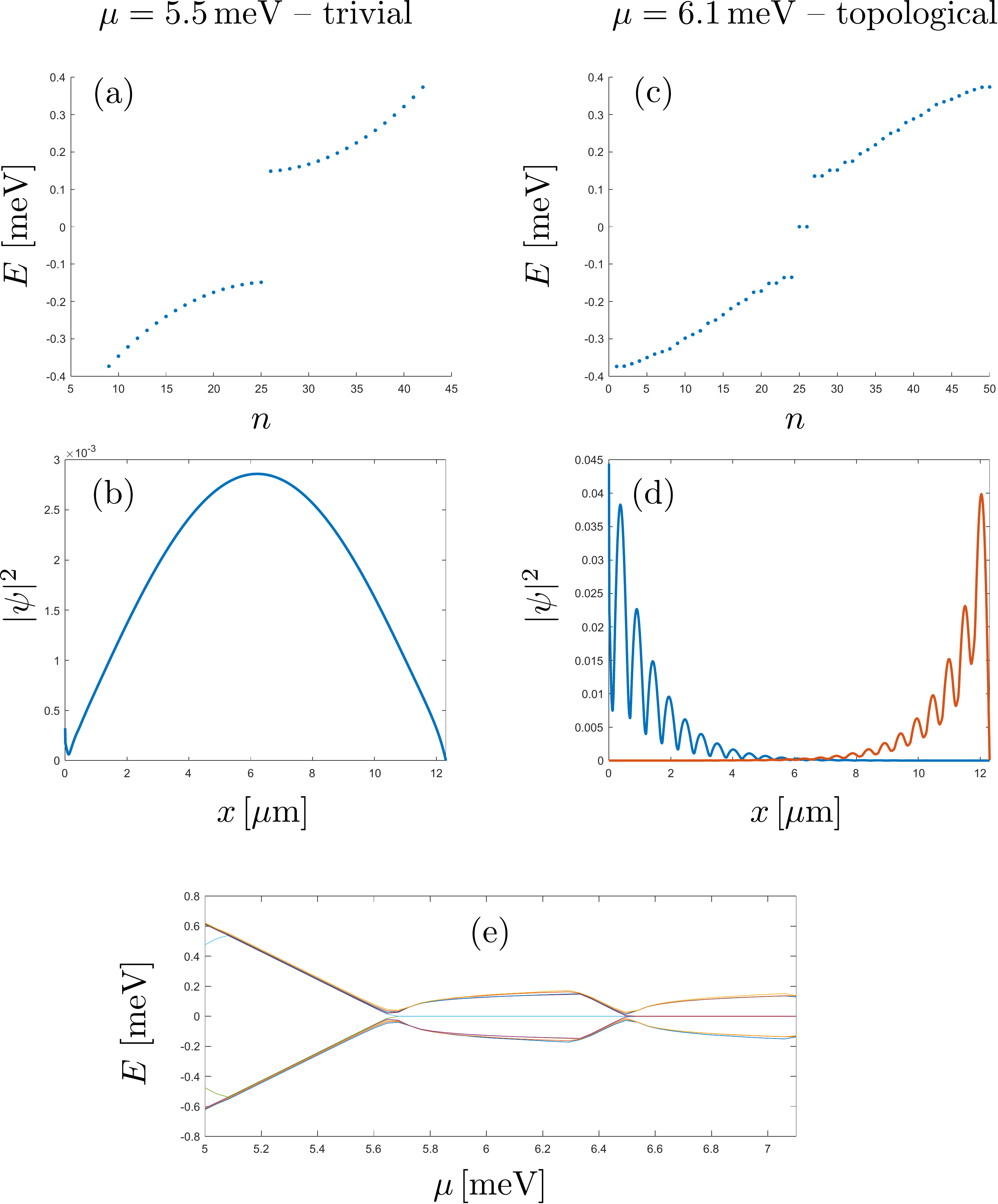}
\par\end{centering}
\caption{\label{fig:real_space_spect_and_wavefcn}Topologically trivial (a-b)
and non-trivial (c-d) phases of the proximitized CNT. The trivial
phase is gapped and all of its eigenstates are bulk states. In contrast,
the topological phase exhibits a pair of zero-energy modes within
the bulk gap, and their wavefunctions are localized at the CNT's edges.
The spectra are shown in (a), (c); in (b), (d) we show the absolute
value squared of the electronic wavefunctions. The two curves in (d)
are obtained as follows: Starting from two degenerate zero-energy
eigenstates, we multiply one of them by a global phase so its first
component matches the other's, and then we take the symmetric and
anti-symmetric combination of the resulting wavefunctions. This yields
two states localized at opposite edges of the CNT; a tunneling density
of states measurement would produce the sum of the two curves displayed.
(e) Ordered eigen-energies (absolute value squared) as a function
of the chemical potential $\mu$. The topological phase transitions
are observed as gap closings and re-openings, where in the topological
phase zero-energy modes persist inside the bulk gap.}
\end{figure}

The phase transition may be observed in Fig.~\ref{fig:real_space_spect_and_wavefcn},
where we show the BdG spectrum and the lowest-energy wavefunction
for the trivial and topological phases (we control the crossover by
tuning the chemical potential $\mu$). The trivial phase is gapped
and has no edge modes, whereas the topological phase exhibits two
zero-energy modes localized at the edges of the CNT. The Majorana
localization length can be roughly estimated as $\xi_{\rm M}=v_{\rm F}/E_{\rm g}$,
which is of the order of $1\mu\text{m}$.

Another way to observe the topological phase transition is inspecting
the BdG spectrum as a function of one of the parameters, e.g. $\mu$,
see Fig.~\ref{fig:real_space_spect_and_wavefcn}e. The topological
phase transitions are signaled by closings of the bulk gap. The gap
then re-opens inside the phases, but in the topological phase, zero-energy
modes clearly appear inside the gap. 

\section {Topological superconductivity without a magnetic flux}\label{triptopssec}

The combination of a tunable CNT and a superconducting TMD substrate
may give rise to topological superconductivity in the absence of any
magnetic field. This is only made possible in the presence of strong
enough electron-electron interactions~\cite{NoGoMAJORANA}, which heavily suppress
the proximity induced spin-singlet component of the superconductivity
as compared to the spin-triplet one. Then, a time-reversal invariant
topological superconducting phase~\cite{TritopsHFpaper,TritopsReview}
manifests itself in the system.

\begin{figure}
\begin{centering}
\includegraphics[scale=0.42]{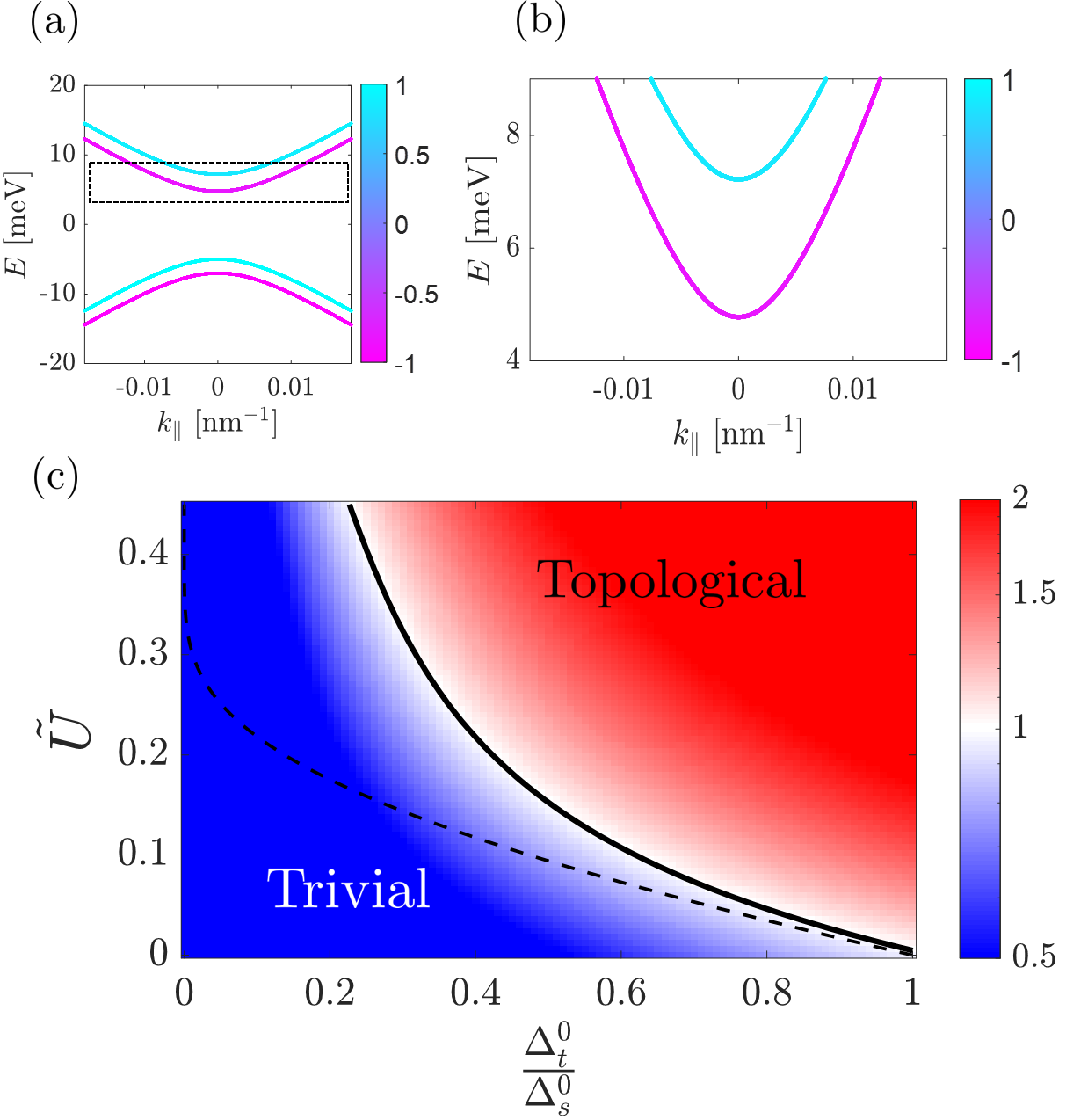}
\par\end{centering}
\caption{\label{fig:phaseDiag}
(a) Normal-state spectrum of the CNT with no magnetic field and with a finite curvature-induced gap.
The color scale indicates $\left\langle \nu_{x}\right\rangle $. All bands are exactly
spin degenerate. Parameters used here are $v_{\rm F}\alpha=0.4$ meV,
$v_{\rm F}\alpha'=0.3$ meV, $V_{\nu}=1$ meV, and $v_{\rm F}\kappa=6$ meV.
(b) Zoom-in on a small part of the spectrum marked by dashed lines
in (a). (c) Phase diagram of the spin-degenerate system with interactions,
in the presence of the two types of proximity induced superconducting
pairing. As a function of initial conditions, we plot the ratio $\frac{\Delta_{t}\left(\ell^{*}\right)}{\Delta_{s}\left(\ell^{*}\right)}$,
with $\ell^{*}$ the RG time at which the first of the pairing potentials reaches strong coupling. The solid black line marks the transition
$\frac{\Delta_{t}\left(\ell^{*}\right)}{\Delta_{s}\left(\ell^{*}\right)}=1$.
The dashed black line marks the same transition, calculated using the approximate phase boundary Eq.~\eqref{eq:convexborder}.
As expected, this approximation overestimates the prominence of the topological sector.
Notice that the color scale here is logarithmic, and that we used the single parameter $\tilde{U}$
to account for all interactions. For this plot, we used the initial condition $\delta_s^0=0.05$. The definition of this parameter, as well as the full RG equations, are all found in Appendix~\ref{app:tritopsApp}.
}
\end{figure}

We now consider thin CNTs, in which interactions play a more significant
role~\cite{CNTinteractionVr}, and that have a substantial curvature-induced
gap, such as zigzag CNTs $(3n,0)$, with an integer $n$. 
The non-interacting part of the Hamiltonian is described by the
low-energy theory
\begin{equation}
H_{0}=v_{\rm F}\left(k\rho_{y}+\left(\kappa+\alpha\nu_{z}\sigma_{z}\right)\rho_{x}+\alpha'\nu_{z}\sigma_{z}\right)+V_{\nu}\nu_{x}-\mu,\label{eq:tritopsH0}
\end{equation}
and $v_{\rm F}\kappa$ plays the role of the curvature gap. When $H_{0}$
is dominated by $\kappa$, the spectrum rather simplifies, see Figs.~\ref{fig:phaseDiag}a,b. Spin degeneracy is not lifted (since no magnetic
flux is introduced), and the different bands are approximate $\nu_{x}$
eigenstates, slightly modified by the presence of spin-orbit-coupling terms. 

By properly adjusting a gate voltage, and thus $\mu$, one can tune
to a point where the Fermi level crosses a single spin-degenerate
band. At this level we effectively describe our system as a 1D system
with two spin species. This system is proximity coupled to a superconductor
having a spin-singlet component $\Delta_{s}$, as well as a spin-triplet component $\Delta_{t}$.

We note that the CNT
origin of this effective Hamiltonian should not be entirely cast away.
For example, upon adding the BdG term $\Delta_{t}^{0}\rho_{y}\sigma_{x}\tau_{x}$ and examing the energy spectrum,  we find that the curvature term $\kappa$ reduces the pairing gap to $\Delta_{t}\approx\Delta_{t}^{0}\sqrt{\frac{k_{F}^{2}}{k_{F}^{2}+\kappa^{2}}}$. 

Upon linearization of the spectrum near the Fermi points, we write
the Hamiltonian density

\begin{align}
{\cal H} & =\sum_{\sigma,r}\psi_{r\sigma}^{\dagger}\left(irv_{\rm F}\partial_{x}-\mu\right)\psi_{r\sigma}+{\cal H}_{{\rm int}}\nonumber \\
 & +\left[\Delta_{s}\left(\psi_{R\uparrow}\psi_{L\downarrow}+\psi_{L\uparrow}\psi_{R\downarrow}\right)+{\rm h.c.}\right]\nonumber \\
 & +\left[\Delta_{t}\left(\psi_{R\uparrow}\psi_{L\uparrow}-\psi_{R\downarrow}\psi_{L\downarrow}\right)+{\rm h.c.}\right],\label{eq:FermionTritops}
\end{align}
where $\psi_{r\sigma}$ annihilates a fermion with spin $\sigma$
and chirality $r=R,L$, and ${\cal H}_{{\rm int}}$ accounts for interactions.
The form of the pairing $\Delta_{t}$ is dictated by the spin polarization of the electrons in the CNT along the tube axis, the Ising nature of the TMD with out-of-plane spin polarization, and  time-reversal
symmetry. The relative minus sign between the two-species spin-triplet proximity term ensures that for $\Delta_{s}=0$ the system is in the
topological phase~\cite{TRITOPSinvariant}. To get a transition into a
trivial phase, the BdG gap must be closed. For the non-interacting Hamiltonian this occurs when $\left|\Delta_{s}\right|=\left|\Delta_{t}\right|$,
hence we have the topological condition
\begin{equation}\label{eq:topocondnonint}
\left|\Delta_{s}\right|<\left|\Delta_{t}\right|.
\end{equation}

One generically expects the singlet proximity component to be greater (even if comparable in size) to the spin-triplet one.
We find, however, that although this may indeed be true for the \textit{bare} values of the superconducting gaps, interactions renormalize both proximity terms. This renormalization naturally favors the triplet over the singlet component, as we show in Appendix~\ref{app:tritopsApp}.

More concretely, we find at the tree level of the renormalization group (RG) flow that the topological condition Eq.~\eqref{eq:topocondnonint} is modified by accounting for interactions into
\begin{equation}
\Delta_{t,0}\geq\left(\Delta_{s,0}\right)^{\frac{4-K_{c}^{-1}-K_{s}^{-1}}{4-K_{c}^{-1}-K_{s}-y}},\label{eq:convexborder}
\end{equation}
with $\Delta_{s/t,0}$ the bare proximity terms, $K_c$ and $K_s$ the Luttinger parameters of the charge and spin sectors, respectively, and $y$ a dimensionless coupling accounting for backscattering interactions (see Appendix~\ref{app:tritopsApp} for more details). 
For generic repulsive interactions, one has the bare values
\[
K_{c}^{0}<1,\,\,\,\,\,K_{s}^{0}>1,\,\,\,\,\,y^{0}>0,
\]
and thus it is becomes clear from Eq.~\eqref{eq:convexborder} that repulsive interactions enhance the topological part of the phase diagram. 
For small Hubbard-like interactions characterized by a single dimensionless parameter $\tilde{U}>0$, Eq.~\eqref{eq:convexborder} can be written in a simpler form, $\Delta_{t,0}\geq\left(\Delta_{s,0}\right)^{\frac{1}{1-2\tilde{U}}}$.
The interaction-dependent phase boundary
gets distorted by the higher-order corrections to the RG flow, which tend to favor the trivial phase.

We integrate the full RG equations derived in Appendix~\ref{app:tritopsApp}, up to the point where one of the pairing potentials reaches strong coupling.
At this point, the ratio between the two
$\Delta$'s is extracted, and it is plotted in Fig.~\ref{fig:phaseDiag}c. 
The main observation is the fact that even if $\Delta_{s}$ is initially significantly
larger than $\Delta_{t}$, as is presumably the case when proximitizing
the system to a superconducting TMD, \textit{strong enough interactions
drive the system to a time-reversal-invariant topological superconducting phase}.

We finally comment on the departure from the Hubbard-like interactions. 
Longer-range interactions will have a three-fold
effect on our equations: they will make $y$ smaller, drive $K_{s}$
much closer to 1, and significantly decrease the value of $K_{c}$.
For a given $K_{c}$ the first two of these effects will obviously
favor the singlet pairing, and reduce the topological area in parameter
space. However, for an even smaller $K_{c}$, there is a possibility
that $\Delta_{s}$ becomes irrelevant, whereas $\Delta_{t}$ is still
(perhaps barely) relevant.
Thus, long-range interactions do not necessarily eliminate the topological phase from the phase diagram, although one does expect to find a smaller topological gap in this case.

\section {Conclusions}\label{Conclus}

In this work we have presented a novel scheme for realizing robust
1D topological superconductivity in an accessible experimental platform
-- CNTs, which are very clean and have true 1D properties. By applying
a magnetic field parallel to the axis of the CNT, and exploiting an
orbital effect instead of Zeeman splitting, we were able to demonstrate
the emergence of the topological phase at relatively low magnetic
fields, compared to previous suggestions~\cite{GrifoniMajoranaCNT}.

The proposed scheme also requires the use of a superconductor with a spin-triplet
component as the reservoir of electron pairs for our system. The use
of superconducting monolayer TMDs, such as $\text{NbSe}_{2}$, is
proposed, due to the unusually strong out-of-plane spin-locking characteristic
of their charge carriers. This in turn ensures that a significant
spin-triplet component exists in the superconducting wavefunction,
which may then couple to the half-metallic CNT. Using monolayer TMDs
allows some control of the chemical potential, and also ensures that
superconductivity is preserved when a moderate magnetic field is applied.
A promising alternative proximitizing substrate is magic-angle twisted bilayer graphene, which was recently discovered to support a superconducting phase~\cite{MITtblg}. It was suggested that significant interaction effects may favor spin-triplet superconductivity in this system~\cite{TBLGtriplet}, making it adequate for our scheme. 

In fact, we have shown that half-metallicity is not strictly necessary, since
a ``one-and-a-half'' metallic state can be made topologically superconducting
due to an odd number (three) of topologically gapped channels. Moreover,
in certain cases a residual spin-singlet component of the superconductor
may help drive an otherwise trivial regime, i.e., where the normal-state spectrum has an even number of channels,
to a topological one, provided that $\Delta_{s}$ is not too large compared to the spin splitting (induced by the magnetic flux and SOC).

In addition, we analyzed the fate of the system in the absence of a magnetic field.
We found that the presence of a spin-triplet component in the superconducting
proximitizing substrate can make a time-reversal-invariant topological phase accessible.
Since interactions tend to decrease the amplitude of
spin-singlet pairing more than spin-triplet pairing, one may end up in a state with a Majorana-Kramers pair of zero modes, protected by time-reversal symmetry.
Our analysis of the Coulomb-interaction effects in the CNT on the induced pairing have important consequences for the finite-flux case as well.
Namely, they suggest a generic suppression of the spin-singlet component as compared to the triplet one. This in turn may enable us to access the topological phase with an even smaller applied magnetic field, see Figs.~\ref{fig:4muphis}a,b.

To examine the real-space wavefunction, the numerical values of the gaps, and the finite-size effects we have simulated numerically a thin $\left(6,0\right)$ CNT, with $1.2\cdot 10^6$ carbon-atom sites, and found a good agreement with the low-energy Hamiltonian Eq.~\eqref{eq:bdgHamiltonian}.
Using the low-energy description, with physical parameters of thicker CNTs (given in Appendix~\ref{app:cheatsheet}) we find that the magnetic field required to tune them into the topological phase is much smaller than the one we simulated.
For example, for a CNT of radius 2.5 nm on top of NbSe$_2$ our consideration gives a topological superconductor at $200$ mT with a gap of order 0.1 meV. A thicker CNT requires an even smaller field to become topological, but the gap is expected to be smaller. 

\begin {acknowledgments}
We thank Erez Berg and Ady Stern for insightful discussions. This work was partially supported by ISF Research Grant in Quantum Technologies and Science.
\end {acknowledgments}

\begin {appendix}

\section {Details of the tight-binding simulations}\label{appendix:TB_details}

Here we provide some technical details regarding the implementation
of the tight-binding model. First, let us explicitly formulate the
Peierls phase~\cite{peierls_zur_1933} associated with the magnetic flux $\Phi$ through the
tube. Labeling the location of site $i$ in the graphene lattice $\mathbf{r}_{i}$,
the Peierls phase is given by~\cite{efroni_topological_2017}
\begin{equation}
A_{ij}=2\pi\frac{\Phi}{\Phi_{0}}\frac{\left(\mathbf{r}_{i}-\mathbf{r}_{j}\right)\cdot\mathbf{C}}{\left|\mathbf{C}\right|^{2}},
\end{equation}
where $\Phi_{0}=h/e$ is the magnetic flux quantum. The SOC terms
are essentially spin-dependent hopping terms, and are thus given by~\cite{efroni_topological_2017}
\begin{equation}
\Delta_{o/z,ij}^{\text{SO}}=\Delta_{o/z}^{\text{SO}}\left(\frac{\left(\mathbf{r}_{i}-\mathbf{r}_{j}\right)\cdot\mathbf{C}_{h}}{\left|\mathbf{r}_{i}-\mathbf{r}_{j}\right|\cdot\left|\mathbf{C}_{h}\right|}\right)^{2}\text{sign}\left(\left(\mathbf{r}_{i}-\mathbf{r}_{j}\right)\cdot\mathbf{C}_{h}\right),
\end{equation}
where $\Delta_{o/z}^{\text{SO}}$ is the ``bare'' orbital / Zeeman
SOC strength.

The inter-valley mixing term $V_{\nu}$ in the low-energy Hamiltonian
Eq.~\eqref{eq:continuumNormalH} corresponds to the angle-dependent
potential $V\left(\theta\right)$ in the tight-binding description
Eq.~\eqref{eq:H_CNT_tight_binding}. To implement this term, we first
calculate the angle of each site along the CNT's circumference $\theta_{i}=2\pi\mathbf{r}_{i}\cdot\mathbf{C}/\left|\mathbf{C}\right|^{2}$.
We examined several forms for the function $V\left(\theta\right)$
which is aimed at mimicking the effect of the gate potential, and
they all yielded similar results. One possible form is a step-like
structure, 
\begin{equation}
V_{\text{step}}\left(\theta\right)=\begin{cases}
V_{0}, & \theta_{1}\leq\theta\leq\theta_{2}\\
0, & \text{otherwise.}
\end{cases}
\end{equation}
Another reasonable form is a Gaussian potential, 
\begin{equation}
V_{\text{Gaussian}}\left(\theta\right)=V_{0}\exp\left[-\frac{\left(\theta-\theta_{0}\right)^{2}}{\left(\Delta\theta\right)^{2}}\right].
\end{equation}
In our simulations we mostly used $V_{\text{step}}\left(\theta\right)$
with $\theta_{1}=0,\theta_{2}=\pi$, i.e. the voltage at half of the
CNT is shifted, thus breaking the azimuthal symmetry.

\section{Experimental parameters }\label{app:cheatsheet}
We bring here for convenience explicit expressions that relate the effective low-energy parameters to experimental parameters of the CNT.
The chiral vector $\mathbf{C}=\left(n,m\right)$ of the CNT is related to its radius by  
\begin{equation}
    R=\frac{a\left|\mathbf{C}\right|}{2\pi},
\end{equation}
with the unit cell size $a=\sqrt{3}a_{\rm{CC}}\approx0.25{\rm nm}$ ($a_{\rm{CC}}$ is the separation between nearest-neighbor carbon atoms), and $\left|\mathbf{C}\right|=\sqrt{n^{2}+nm+m^{2}}$.

The magnetic flux term $v_{\rm F}\phi= \hbar v_{\rm F} \frac{1}{R} \frac{\Phi}{\Phi_0}$, with $\Phi_0=2\cdot 10^{-15}$ T$\cdot$m$^2$ the flux quantum, and $\Phi=\pi R^2 B_z$ the flux through the CNT. The Fermi velocity of the Dirac cones is estimated as $v_{\rm F}\approx 10^6\,\frac{\rm m}{\rm sec}$. We thus find
\begin{equation}
    v_{\rm F} \phi \approx B_z \left[{\rm T}\right] R\left[{\rm nm}\right]{\rm meV}\approx \frac{\left|\mathbf{C}\right|}{24}B_z \left[{\rm T}\right]{\rm meV}.
\end{equation}

The spin-orbit-coupling term $v_{\rm F} \alpha$ can be estimated from previous studies~\cite{IlaniSOC} which found the spin-orbit gap $\Delta_{{\rm SO}}\approx0.4\,{\rm meV}$ for CNT with radius of 2.5 nm, and thus
\begin{equation}
v_{\rm F} \alpha \approx 1\frac{\rm meV}{R\left[{\rm nm}\right]}\approx\frac{25}{\left|\mathbf{C}\right|}\,{\rm meV}.
\end{equation}
For reference, we bring here also the Zeeman energy $V_{\rm Z}=g\mu_{\rm B} B_{z}$, with $g\approx2$ and $\mu_{\rm B}$ the Bohr magneton,
\begin{equation}
V_{\rm Z}  \approx0.11{B\left[{\rm T}\right]}\,{\rm meV}.
\end{equation}

\section{Interactions in the time-reversal invariant case}\label{app:tritopsApp}
To account for the important effect of electron-electron interactions introduced in Sec.~\ref{triptopssec},
we bosonize the Hamiltonian Eq.~\eqref{eq:FermionTritops} using standard
identities, $\psi_{r\sigma}\sim\frac{1}{\sqrt{2\pi a}}e^{-i\left(r\phi_{\sigma}-\theta_{\sigma}\right)}$,
with $a$ some short-distance cutoff, and the bosonic fields satisfying the algebra $\left[\phi_{\sigma}\left(x\right),\partial_{x}\theta_{\sigma'}\left(x'\right)\right]=i\pi\delta_{\alpha\beta}\delta\left(x-x'\right)$~\cite{giamarchi2004quantum}.
By defining the charge and spin sectors $\phi_{c,s}\equiv\frac{\phi_{\uparrow}\pm\phi_{\downarrow}}{\sqrt{2}}$,
this representation allows us to re-organize our Hamiltonian into
four parts, ${\cal H}={\cal H}_{c}+{\cal H}_{s}+{\cal H}_{\Delta_{s}}+{\cal H}_{\Delta_{t}}$,
with \begin {subequations}
\begin{equation}
{\cal H}_{c}=\frac{v_{c}}{2\pi}\left[K_{c}^{-1}\left(\partial_{x}\phi_{c}\right)^{2}+K_{c}\left(\partial_{x}\theta_{c}\right)^{2}\right],\label{eq:Hcboson}
\end{equation}
\begin{align}
{\cal H}_{s}&=\frac{v_{s}}{2\pi}\left[K_{s}^{-1}\left(\partial_{x}\phi_{s}\right)^{2}+K_{s}\left(\partial_{x}\theta_{s}\right)^{2}\right]\nonumber\\ 
& + \frac{g}{2\pi^{2}a^{2}}\cos\left(\sqrt{8}\phi_{s}\right),
\end{align}
\begin{equation}
{\cal H}_{\Delta_{s}}=\frac{2\Delta_{s}}{\pi a}\cos\left(\sqrt{2}\theta_{c}\right)\cos\left(\sqrt{2}\phi_{s}\right),
\end{equation}
\begin{equation}
{\cal H}_{\Delta_{t}}=\frac{2\Delta_{t}}{\pi a}\cos\left(\sqrt{2}\theta_{c}\right)\cos\left(\sqrt{2}\theta_{s}\right),\label{eq:HdtBoson}
\end{equation}
\end {subequations} with the interaction incorporated into the so-called
Luttinger parameters $K_{\eta},v_{\eta}$, ($\eta=c,s$) and into the backscattering
term $g$ \footnote{$g$ is the usual backscattering $g_{1\perp}$ term in the standard
g-ology language.}. For our purposes, we will approximate $v_c\approx v_s\equiv v$.
This representation makes clear the competition between the two pairing
terms, as well as between the backscattering $g$ and $\Delta_{s}$,
whose energy cannot be simultaneously minimized for any $g>0$. 

Defining the dimensionless constants $y=\frac{g}{\pi v},\delta_{s/t}=\frac{4\Delta_{s/t}a}{v}$,
the RG equations may be derived in a straightforward manner~\cite{cardy_1996},\begin {subequations}
\begin{equation}
\frac{d}{d\ell}y=\left(2-2K_{s}\right)y-\frac{1}{4}\delta_{s}^{2},\label{eq:RG1}
\end{equation}
\begin{equation}
\frac{d}{d\ell}\delta_{s}=\frac{1}{2}\left(4-K_{c}^{-1}-K_{s}-y\right)\delta_{s},\label{eq:dsRG}
\end{equation}
\begin{equation}
\frac{d}{d\ell}\delta_{t}=\frac{1}{2}\left(4-K_{c}^{-1}-K_{s}^{-1}\right)\delta_{t},\label{eq:DTrg}
\end{equation}
\begin{equation}
\frac{d}{d\ell}K_{s}=-\frac{1}{2}K_{s}^{2}\left(y^{2}+\frac{1}{4}\delta_{s}^{2}\right)+\frac{1}{8}\delta_{t}^{2},
\end{equation}
\begin{equation}
\frac{d}{d\ell}K_{c}=\frac{1}{8}\left(\delta_{t}^{2}+\delta_{s}^{2}\right).\label{eq:RG2}
\end{equation}
\end{subequations} At the tree level, one may consider the
RG flow under Eqs.~\eqref{eq:dsRG}--\eqref{eq:DTrg} only, yielding the condition Eq.~\eqref{eq:convexborder}. 
Importantly, in the presence of repulsive interactions, this shows that $\Delta_{t}$ is \emph{always} more relevant than $\Delta_{s}$.
However, their initial values and their respective distances from strong coupling will determine the nature of the pairing in the low-energy limit. 

Assuming Hubbard-like interactions, we may approximate
\begin{equation}
K_{c}\approx\sqrt{\frac{1+\frac{g}{2\pi v}}{1+\frac{3g}{2\pi v}}},\,\,\,\,\,K_{s}\approx\sqrt{\frac{1+\frac{g}{2\pi v}}{1-\frac{g}{2\pi v}}},\label{eq:Hubbardform}
\end{equation}
and we also define $\tilde{U}=\frac{g}{2\pi v}$. This form is
reasonable, as in an experimental setup which includes metallic gates
and bulk superconductors, the interactions are fairly well-screened.
We note that for small $\tilde{U}$, we may expand the above coefficients
and approximate the previous phase boundary in a more manageable form,
$\Delta_{t,0}\geq\left(\Delta_{s,0}\right)^{\frac{1}{1-2\tilde{U}}}$.

\end {appendix}

\bibliography{refBibtex}

\begin{thebibliography}{52}%
\makeatletter
\providecommand \@ifxundefined [1]{%
 \@ifx{#1\undefined}
}%
\providecommand \@ifnum [1]{%
 \ifnum #1\expandafter \@firstoftwo
 \else \expandafter \@secondoftwo
 \fi
}%
\providecommand \@ifx [1]{%
 \ifx #1\expandafter \@firstoftwo
 \else \expandafter \@secondoftwo
 \fi
}%
\providecommand \natexlab [1]{#1}%
\providecommand \enquote  [1]{``#1''}%
\providecommand \bibnamefont  [1]{#1}%
\providecommand \bibfnamefont [1]{#1}%
\providecommand \citenamefont [1]{#1}%
\providecommand \href@noop [0]{\@secondoftwo}%
\providecommand \href [0]{\begingroup \@sanitize@url \@href}%
\providecommand \@href[1]{\@@startlink{#1}\@@href}%
\providecommand \@@href[1]{\endgroup#1\@@endlink}%
\providecommand \@sanitize@url [0]{\catcode `\\12\catcode `\$12\catcode
  `\&12\catcode `\#12\catcode `\^12\catcode `\_12\catcode `\%12\relax}%
\providecommand \@@startlink[1]{}%
\providecommand \@@endlink[0]{}%
\providecommand \url  [0]{\begingroup\@sanitize@url \@url }%
\providecommand \@url [1]{\endgroup\@href {#1}{\urlprefix }}%
\providecommand \urlprefix  [0]{URL }%
\providecommand \Eprint [0]{\href }%
\providecommand \doibase [0]{http://dx.doi.org/}%
\providecommand \selectlanguage [0]{\@gobble}%
\providecommand \bibinfo  [0]{\@secondoftwo}%
\providecommand \bibfield  [0]{\@secondoftwo}%
\providecommand \translation [1]{[#1]}%
\providecommand \BibitemOpen [0]{}%
\providecommand \bibitemStop [0]{}%
\providecommand \bibitemNoStop [0]{.\EOS\space}%
\providecommand \EOS [0]{\spacefactor3000\relax}%
\providecommand \BibitemShut  [1]{\csname bibitem#1\endcsname}%
\let\auto@bib@innerbib\@empty
\bibitem [{\citenamefont {Read}\ and\ \citenamefont {Green}(2000)}]{ReadGreen}%
  \BibitemOpen
  \bibfield  {author} {\bibinfo {author} {\bibfnamefont {N.}~\bibnamefont
  {Read}}\ and\ \bibinfo {author} {\bibfnamefont {D.}~\bibnamefont {Green}},\
  }\href {\doibase 10.1103/PhysRevB.61.10267} {\bibfield  {journal} {\bibinfo
  {journal} {Phys. Rev. B}\ }\textbf {\bibinfo {volume} {61}},\ \bibinfo
  {pages} {10267} (\bibinfo {year} {2000})}\BibitemShut {NoStop}%
\bibitem [{\citenamefont {Kitaev}(2001)}]{Kitaev_2001}%
  \BibitemOpen
  \bibfield  {author} {\bibinfo {author} {\bibfnamefont {A.~Y.}\ \bibnamefont
  {Kitaev}},\ }\href {\doibase 10.1070/1063-7869/44/10s/s29} {\bibfield
  {journal} {\bibinfo  {journal} {Physics-Uspekhi}\ }\textbf {\bibinfo {volume}
  {44}},\ \bibinfo {pages} {131} (\bibinfo {year} {2001})}\BibitemShut
  {NoStop}%
\bibitem [{\citenamefont {Lutchyn}\ \emph {et~al.}(2010)\citenamefont
  {Lutchyn}, \citenamefont {Sau},\ and\ \citenamefont
  {Das~Sarma}}]{LutchynMajorana}%
  \BibitemOpen
  \bibfield  {author} {\bibinfo {author} {\bibfnamefont {R.~M.}\ \bibnamefont
  {Lutchyn}}, \bibinfo {author} {\bibfnamefont {J.~D.}\ \bibnamefont {Sau}}, \
  and\ \bibinfo {author} {\bibfnamefont {S.}~\bibnamefont {Das~Sarma}},\ }\href
  {\doibase 10.1103/PhysRevLett.105.077001} {\bibfield  {journal} {\bibinfo
  {journal} {Phys. Rev. Lett.}\ }\textbf {\bibinfo {volume} {105}},\ \bibinfo
  {pages} {077001} (\bibinfo {year} {2010})}\BibitemShut {NoStop}%
\bibitem [{\citenamefont {Oreg}\ \emph {et~al.}(2010)\citenamefont {Oreg},
  \citenamefont {Refael},\ and\ \citenamefont {von Oppen}}]{OregMajoran}%
  \BibitemOpen
  \bibfield  {author} {\bibinfo {author} {\bibfnamefont {Y.}~\bibnamefont
  {Oreg}}, \bibinfo {author} {\bibfnamefont {G.}~\bibnamefont {Refael}}, \ and\
  \bibinfo {author} {\bibfnamefont {F.}~\bibnamefont {von Oppen}},\ }\href
  {\doibase 10.1103/PhysRevLett.105.177002} {\bibfield  {journal} {\bibinfo
  {journal} {Phys. Rev. Lett.}\ }\textbf {\bibinfo {volume} {105}},\ \bibinfo
  {pages} {177002} (\bibinfo {year} {2010})}\BibitemShut {NoStop}%
\bibitem [{\citenamefont {Kitaev}(2003)}]{KITAEV20032}%
  \BibitemOpen
  \bibfield  {author} {\bibinfo {author} {\bibfnamefont {A.}~\bibnamefont
  {Kitaev}},\ }\href {\doibase https://doi.org/10.1016/S0003-4916(02)00018-0}
  {\bibfield  {journal} {\bibinfo  {journal} {Annals of Physics}\ }\textbf
  {\bibinfo {volume} {303}},\ \bibinfo {pages} {2 } (\bibinfo {year}
  {2003})}\BibitemShut {NoStop}%
\bibitem [{\citenamefont {Nayak}\ \emph {et~al.}(2008)\citenamefont {Nayak},
  \citenamefont {Simon}, \citenamefont {Stern}, \citenamefont {Freedman},\ and\
  \citenamefont {Das~Sarma}}]{NayakStern}%
  \BibitemOpen
  \bibfield  {author} {\bibinfo {author} {\bibfnamefont {C.}~\bibnamefont
  {Nayak}}, \bibinfo {author} {\bibfnamefont {S.~H.}\ \bibnamefont {Simon}},
  \bibinfo {author} {\bibfnamefont {A.}~\bibnamefont {Stern}}, \bibinfo
  {author} {\bibfnamefont {M.}~\bibnamefont {Freedman}}, \ and\ \bibinfo
  {author} {\bibfnamefont {S.}~\bibnamefont {Das~Sarma}},\ }\href {\doibase
  10.1103/RevModPhys.80.1083} {\bibfield  {journal} {\bibinfo  {journal} {Rev.
  Mod. Phys.}\ }\textbf {\bibinfo {volume} {80}},\ \bibinfo {pages} {1083}
  (\bibinfo {year} {2008})}\BibitemShut {NoStop}%
\bibitem [{\citenamefont {Deng}\ \emph {et~al.}(2012)\citenamefont {Deng},
  \citenamefont {Yu}, \citenamefont {Huang}, \citenamefont {Larsson},
  \citenamefont {Caroff},\ and\ \citenamefont {Xu}}]{Deng2012}%
  \BibitemOpen
  \bibfield  {author} {\bibinfo {author} {\bibfnamefont {M.~T.}\ \bibnamefont
  {Deng}}, \bibinfo {author} {\bibfnamefont {C.~L.}\ \bibnamefont {Yu}},
  \bibinfo {author} {\bibfnamefont {G.~Y.}\ \bibnamefont {Huang}}, \bibinfo
  {author} {\bibfnamefont {M.}~\bibnamefont {Larsson}}, \bibinfo {author}
  {\bibfnamefont {P.}~\bibnamefont {Caroff}}, \ and\ \bibinfo {author}
  {\bibfnamefont {H.~Q.}\ \bibnamefont {Xu}},\ }\href {\doibase
  10.1021/nl303758w} {\bibfield  {journal} {\bibinfo  {journal} {Nano Letters}\
  }\textbf {\bibinfo {volume} {12}},\ \bibinfo {pages} {6414} (\bibinfo {year}
  {2012})}\BibitemShut {NoStop}%
\bibitem [{\citenamefont {Mourik}\ \emph {et~al.}(2012)\citenamefont {Mourik},
  \citenamefont {Zuo}, \citenamefont {Frolov}, \citenamefont {Plissard},
  \citenamefont {Bakkers},\ and\ \citenamefont {Kouwenhoven}}]{Mourik1003}%
  \BibitemOpen
  \bibfield  {author} {\bibinfo {author} {\bibfnamefont {V.}~\bibnamefont
  {Mourik}}, \bibinfo {author} {\bibfnamefont {K.}~\bibnamefont {Zuo}},
  \bibinfo {author} {\bibfnamefont {S.~M.}\ \bibnamefont {Frolov}}, \bibinfo
  {author} {\bibfnamefont {S.~R.}\ \bibnamefont {Plissard}}, \bibinfo {author}
  {\bibfnamefont {E.~P. A.~M.}\ \bibnamefont {Bakkers}}, \ and\ \bibinfo
  {author} {\bibfnamefont {L.~P.}\ \bibnamefont {Kouwenhoven}},\ }\href
  {\doibase 10.1126/science.1222360} {\bibfield  {journal} {\bibinfo  {journal}
  {Science}\ }\textbf {\bibinfo {volume} {336}},\ \bibinfo {pages} {1003}
  (\bibinfo {year} {2012})}\BibitemShut {NoStop}%
\bibitem [{\citenamefont {Das}\ \emph {et~al.}(2012)\citenamefont {Das},
  \citenamefont {Ronen}, \citenamefont {Most}, \citenamefont {Oreg},
  \citenamefont {Heiblum},\ and\ \citenamefont {Shtrikman}}]{Das2012}%
  \BibitemOpen
  \bibfield  {author} {\bibinfo {author} {\bibfnamefont {A.}~\bibnamefont
  {Das}}, \bibinfo {author} {\bibfnamefont {Y.}~\bibnamefont {Ronen}}, \bibinfo
  {author} {\bibfnamefont {Y.}~\bibnamefont {Most}}, \bibinfo {author}
  {\bibfnamefont {Y.}~\bibnamefont {Oreg}}, \bibinfo {author} {\bibfnamefont
  {M.}~\bibnamefont {Heiblum}}, \ and\ \bibinfo {author} {\bibfnamefont
  {H.}~\bibnamefont {Shtrikman}},\ }\href {https://doi.org/10.1038/nphys2479}
  {\bibfield  {journal} {\bibinfo  {journal} {Nature Physics}\ }\textbf
  {\bibinfo {volume} {8}},\ \bibinfo {pages} {887 EP } (\bibinfo {year}
  {2012})},\ \bibinfo {note} {article}\BibitemShut {NoStop}%
\bibitem [{\citenamefont {Deng}\ \emph {et~al.}(2016)\citenamefont {Deng},
  \citenamefont {Vaitiekenas}, \citenamefont {Hansen}, \citenamefont {Danon},
  \citenamefont {Leijnse}, \citenamefont {Flensberg}, \citenamefont {Nyg{\r
  a}rd}, \citenamefont {Krogstrup},\ and\ \citenamefont
  {Marcus}}]{MarcusMajoranas}%
  \BibitemOpen
  \bibfield  {author} {\bibinfo {author} {\bibfnamefont {M.~T.}\ \bibnamefont
  {Deng}}, \bibinfo {author} {\bibfnamefont {S.}~\bibnamefont {Vaitiekenas}},
  \bibinfo {author} {\bibfnamefont {E.~B.}\ \bibnamefont {Hansen}}, \bibinfo
  {author} {\bibfnamefont {J.}~\bibnamefont {Danon}}, \bibinfo {author}
  {\bibfnamefont {M.}~\bibnamefont {Leijnse}}, \bibinfo {author} {\bibfnamefont
  {K.}~\bibnamefont {Flensberg}}, \bibinfo {author} {\bibfnamefont
  {J.}~\bibnamefont {Nyg{\r a}rd}}, \bibinfo {author} {\bibfnamefont
  {P.}~\bibnamefont {Krogstrup}}, \ and\ \bibinfo {author} {\bibfnamefont
  {C.~M.}\ \bibnamefont {Marcus}},\ }\href {\doibase 10.1126/science.aaf3961}
  {\bibfield  {journal} {\bibinfo  {journal} {Science}\ }\textbf {\bibinfo
  {volume} {354}},\ \bibinfo {pages} {1557} (\bibinfo {year}
  {2016})}\BibitemShut {NoStop}%
\bibitem [{\citenamefont {Lutchyn}\ \emph {et~al.}(2018)\citenamefont
  {Lutchyn}, \citenamefont {Bakkers}, \citenamefont {Kouwenhoven},
  \citenamefont {Krogstrup}, \citenamefont {Marcus},\ and\ \citenamefont
  {Oreg}}]{Lutchyn2018Review}%
  \BibitemOpen
  \bibfield  {author} {\bibinfo {author} {\bibfnamefont {R.}~\bibnamefont
  {Lutchyn}}, \bibinfo {author} {\bibfnamefont {E.}~\bibnamefont {Bakkers}},
  \bibinfo {author} {\bibfnamefont {L.}~\bibnamefont {Kouwenhoven}}, \bibinfo
  {author} {\bibfnamefont {P.}~\bibnamefont {Krogstrup}}, \bibinfo {author}
  {\bibfnamefont {C.}~\bibnamefont {Marcus}}, \ and\ \bibinfo {author}
  {\bibfnamefont {Y.}~\bibnamefont {Oreg}},\ }\href {\doibase
  10.1038/s41578-018-0003-1} {\bibfield  {journal} {\bibinfo  {journal} {Nature
  Reviews. Materials}\ }\textbf {\bibinfo {volume} {3}},\ \bibinfo {pages} {52}
  (\bibinfo {year} {2018})}\BibitemShut {NoStop}%
\bibitem [{\citenamefont {Iijima}(1991)}]{Iijima1991}%
  \BibitemOpen
  \bibfield  {author} {\bibinfo {author} {\bibfnamefont {S.}~\bibnamefont
  {Iijima}},\ }\href {\doibase 10.1038/354056a0} {\bibfield  {journal}
  {\bibinfo  {journal} {Nature}\ }\textbf {\bibinfo {volume} {354}},\ \bibinfo
  {pages} {56} (\bibinfo {year} {1991})}\BibitemShut {NoStop}%
\bibitem [{\citenamefont {Charlier}\ \emph {et~al.}(2007)\citenamefont
  {Charlier}, \citenamefont {Blase},\ and\ \citenamefont {Roche}}]{RMPcnt}%
  \BibitemOpen
  \bibfield  {author} {\bibinfo {author} {\bibfnamefont {J.-C.}\ \bibnamefont
  {Charlier}}, \bibinfo {author} {\bibfnamefont {X.}~\bibnamefont {Blase}}, \
  and\ \bibinfo {author} {\bibfnamefont {S.}~\bibnamefont {Roche}},\ }\href
  {\doibase 10.1103/RevModPhys.79.677} {\bibfield  {journal} {\bibinfo
  {journal} {Rev. Mod. Phys.}\ }\textbf {\bibinfo {volume} {79}},\ \bibinfo
  {pages} {677} (\bibinfo {year} {2007})}\BibitemShut {NoStop}%
\bibitem [{\citenamefont {Dresselhaus}\ \emph {et~al.}(1995)\citenamefont
  {Dresselhaus}, \citenamefont {Dresselhaus},\ and\ \citenamefont
  {Saito}}]{dresselhaus_physics_1995}%
  \BibitemOpen
  \bibfield  {author} {\bibinfo {author} {\bibfnamefont {M.~S.}\ \bibnamefont
  {Dresselhaus}}, \bibinfo {author} {\bibfnamefont {G.}~\bibnamefont
  {Dresselhaus}}, \ and\ \bibinfo {author} {\bibfnamefont {R.}~\bibnamefont
  {Saito}},\ }\href {\doibase 10.1016/0008-6223(95)00017-8} {\bibfield
  {journal} {\bibinfo  {journal} {Carbon}\ }\bibinfo {series} {Nanotubes},\
  \textbf {\bibinfo {volume} {33}},\ \bibinfo {pages} {883} (\bibinfo {year}
  {1995})}\BibitemShut {NoStop}%
\bibitem [{\citenamefont {Laird}\ \emph {et~al.}(2015)\citenamefont {Laird},
  \citenamefont {Kuemmeth}, \citenamefont {Steele}, \citenamefont
  {Grove-Rasmussen}, \citenamefont {Nyg\aa{}rd}, \citenamefont {Flensberg},\
  and\ \citenamefont {Kouwenhoven}}]{RMPcntcleantransport}%
  \BibitemOpen
  \bibfield  {author} {\bibinfo {author} {\bibfnamefont {E.~A.}\ \bibnamefont
  {Laird}}, \bibinfo {author} {\bibfnamefont {F.}~\bibnamefont {Kuemmeth}},
  \bibinfo {author} {\bibfnamefont {G.~A.}\ \bibnamefont {Steele}}, \bibinfo
  {author} {\bibfnamefont {K.}~\bibnamefont {Grove-Rasmussen}}, \bibinfo
  {author} {\bibfnamefont {J.}~\bibnamefont {Nyg\aa{}rd}}, \bibinfo {author}
  {\bibfnamefont {K.}~\bibnamefont {Flensberg}}, \ and\ \bibinfo {author}
  {\bibfnamefont {L.~P.}\ \bibnamefont {Kouwenhoven}},\ }\href {\doibase
  10.1103/RevModPhys.87.703} {\bibfield  {journal} {\bibinfo  {journal} {Rev.
  Mod. Phys.}\ }\textbf {\bibinfo {volume} {87}},\ \bibinfo {pages} {703}
  (\bibinfo {year} {2015})}\BibitemShut {NoStop}%
\bibitem [{\citenamefont {Sau}\ and\ \citenamefont
  {Tewari}(2013)}]{SauMajoranCNT}%
  \BibitemOpen
  \bibfield  {author} {\bibinfo {author} {\bibfnamefont {J.~D.}\ \bibnamefont
  {Sau}}\ and\ \bibinfo {author} {\bibfnamefont {S.}~\bibnamefont {Tewari}},\
  }\href {\doibase 10.1103/PhysRevB.88.054503} {\bibfield  {journal} {\bibinfo
  {journal} {Phys. Rev. B}\ }\textbf {\bibinfo {volume} {88}},\ \bibinfo
  {pages} {054503} (\bibinfo {year} {2013})}\BibitemShut {NoStop}%
\bibitem [{\citenamefont {Marganska}\ \emph {et~al.}(2018)\citenamefont
  {Marganska}, \citenamefont {Milz}, \citenamefont {Izumida}, \citenamefont
  {Strunk},\ and\ \citenamefont {Grifoni}}]{GrifoniMajoranaCNT}%
  \BibitemOpen
  \bibfield  {author} {\bibinfo {author} {\bibfnamefont {M.}~\bibnamefont
  {Marganska}}, \bibinfo {author} {\bibfnamefont {L.}~\bibnamefont {Milz}},
  \bibinfo {author} {\bibfnamefont {W.}~\bibnamefont {Izumida}}, \bibinfo
  {author} {\bibfnamefont {C.}~\bibnamefont {Strunk}}, \ and\ \bibinfo {author}
  {\bibfnamefont {M.}~\bibnamefont {Grifoni}},\ }\href {\doibase
  10.1103/PhysRevB.97.075141} {\bibfield  {journal} {\bibinfo  {journal} {Phys.
  Rev. B}\ }\textbf {\bibinfo {volume} {97}},\ \bibinfo {pages} {075141}
  (\bibinfo {year} {2018})}\BibitemShut {NoStop}%
\bibitem [{\citenamefont {Klinovaja}\ \emph {et~al.}(2012)\citenamefont
  {Klinovaja}, \citenamefont {Gangadharaiah},\ and\ \citenamefont
  {Loss}}]{CNTmajoranaLoss}%
  \BibitemOpen
  \bibfield  {author} {\bibinfo {author} {\bibfnamefont {J.}~\bibnamefont
  {Klinovaja}}, \bibinfo {author} {\bibfnamefont {S.}~\bibnamefont
  {Gangadharaiah}}, \ and\ \bibinfo {author} {\bibfnamefont {D.}~\bibnamefont
  {Loss}},\ }\href {\doibase 10.1103/PhysRevLett.108.196804} {\bibfield
  {journal} {\bibinfo  {journal} {Phys. Rev. Lett.}\ }\textbf {\bibinfo
  {volume} {108}},\ \bibinfo {pages} {196804} (\bibinfo {year}
  {2012})}\BibitemShut {NoStop}%
\bibitem [{\citenamefont {Qi}\ \emph {et~al.}(2010)\citenamefont {Qi},
  \citenamefont {Hughes},\ and\ \citenamefont {Zhang}}]{TRITOPSinvariant}%
  \BibitemOpen
  \bibfield  {author} {\bibinfo {author} {\bibfnamefont {X.-L.}\ \bibnamefont
  {Qi}}, \bibinfo {author} {\bibfnamefont {T.~L.}\ \bibnamefont {Hughes}}, \
  and\ \bibinfo {author} {\bibfnamefont {S.-C.}\ \bibnamefont {Zhang}},\ }\href
  {\doibase 10.1103/PhysRevB.81.134508} {\bibfield  {journal} {\bibinfo
  {journal} {Phys. Rev. B}\ }\textbf {\bibinfo {volume} {81}},\ \bibinfo
  {pages} {134508} (\bibinfo {year} {2010})}\BibitemShut {NoStop}%
\bibitem [{\citenamefont {Haim}\ and\ \citenamefont
  {Oreg}(2018)}]{TritopsReview}%
  \BibitemOpen
  \bibfield  {author} {\bibinfo {author} {\bibfnamefont {A.}~\bibnamefont
  {Haim}}\ and\ \bibinfo {author} {\bibfnamefont {Y.}~\bibnamefont {Oreg}},\
  }\href@noop {} {\enquote {\bibinfo {title} {Time-reversal-invariant
  topological superconductivity},}\ } (\bibinfo {year} {2018}),\ \Eprint
  {http://arxiv.org/abs/arXiv:1809.06863} {arXiv:1809.06863} \BibitemShut
  {NoStop}%
\bibitem [{\citenamefont {Dresselhaus}\ \emph {et~al.}(1998)\citenamefont
  {Dresselhaus}, \citenamefont {Riichiro} \emph
  {et~al.}}]{dresselhaus1998physical}%
  \BibitemOpen
  \bibfield  {author} {\bibinfo {author} {\bibfnamefont {G.}~\bibnamefont
  {Dresselhaus}}, \bibinfo {author} {\bibfnamefont {S.}~\bibnamefont
  {Riichiro}},  \emph {et~al.},\ }\href@noop {} {\emph {\bibinfo {title}
  {Physical properties of carbon nanotubes}}}\ (\bibinfo  {publisher} {World
  scientific},\ \bibinfo {year} {1998})\BibitemShut {NoStop}%
\bibitem [{\citenamefont {Peierls}(1933)}]{peierls_zur_1933}%
  \BibitemOpen
  \bibfield  {author} {\bibinfo {author} {\bibfnamefont {R.}~\bibnamefont
  {Peierls}},\ }\href {\doibase 10.1007/BF01342591} {\bibfield  {journal}
  {\bibinfo  {journal} {Zeitschrift f{\"u}r Physik}\ }\textbf {\bibinfo
  {volume} {80}},\ \bibinfo {pages} {763} (\bibinfo {year} {1933})}\BibitemShut
  {NoStop}%
\bibitem [{\citenamefont {Ando}(2000)}]{AndoSOC}%
  \BibitemOpen
  \bibfield  {author} {\bibinfo {author} {\bibfnamefont {T.}~\bibnamefont
  {Ando}},\ }\href {\doibase 10.1143/JPSJ.69.1757} {\bibfield  {journal}
  {\bibinfo  {journal} {Journal of the Physical Society of Japan}\ }\textbf
  {\bibinfo {volume} {69}},\ \bibinfo {pages} {1757} (\bibinfo {year}
  {2000})},\ \Eprint
  {http://arxiv.org/abs/https://doi.org/10.1143/JPSJ.69.1757}
  {https://doi.org/10.1143/JPSJ.69.1757} \BibitemShut {NoStop}%
\bibitem [{\citenamefont {Huertas-Hernando}\ \emph {et~al.}(2006)\citenamefont
  {Huertas-Hernando}, \citenamefont {Guinea},\ and\ \citenamefont
  {Brataas}}]{BrataasSOC}%
  \BibitemOpen
  \bibfield  {author} {\bibinfo {author} {\bibfnamefont {D.}~\bibnamefont
  {Huertas-Hernando}}, \bibinfo {author} {\bibfnamefont {F.}~\bibnamefont
  {Guinea}}, \ and\ \bibinfo {author} {\bibfnamefont {A.}~\bibnamefont
  {Brataas}},\ }\href {\doibase 10.1103/PhysRevB.74.155426} {\bibfield
  {journal} {\bibinfo  {journal} {Phys. Rev. B}\ }\textbf {\bibinfo {volume}
  {74}},\ \bibinfo {pages} {155426} (\bibinfo {year} {2006})}\BibitemShut
  {NoStop}%
\bibitem [{\citenamefont {Kuemmeth}\ \emph {et~al.}(2008)\citenamefont
  {Kuemmeth}, \citenamefont {Ilani}, \citenamefont {Ralph},\ and\ \citenamefont
  {McEuen}}]{IlaniSOC}%
  \BibitemOpen
  \bibfield  {author} {\bibinfo {author} {\bibfnamefont {F.}~\bibnamefont
  {Kuemmeth}}, \bibinfo {author} {\bibfnamefont {S.}~\bibnamefont {Ilani}},
  \bibinfo {author} {\bibfnamefont {D.~C.}\ \bibnamefont {Ralph}}, \ and\
  \bibinfo {author} {\bibfnamefont {P.~L.}\ \bibnamefont {McEuen}},\ }\href
  {https://doi.org/10.1038/nature06822} {\bibfield  {journal} {\bibinfo
  {journal} {Nature}\ }\textbf {\bibinfo {volume} {452}},\ \bibinfo {pages}
  {448 EP } (\bibinfo {year} {2008})}\BibitemShut {NoStop}%
\bibitem [{\citenamefont {Izumida}\ \emph {et~al.}(2009)\citenamefont
  {Izumida}, \citenamefont {Sato},\ and\ \citenamefont {Saito}}]{alphaTag}%
  \BibitemOpen
  \bibfield  {author} {\bibinfo {author} {\bibfnamefont {W.}~\bibnamefont
  {Izumida}}, \bibinfo {author} {\bibfnamefont {K.}~\bibnamefont {Sato}}, \
  and\ \bibinfo {author} {\bibfnamefont {R.}~\bibnamefont {Saito}},\ }\href
  {\doibase 10.1143/JPSJ.78.074707} {\bibfield  {journal} {\bibinfo  {journal}
  {Journal of the Physical Society of Japan}\ }\textbf {\bibinfo {volume}
  {78}},\ \bibinfo {pages} {074707} (\bibinfo {year} {2009})},\ \Eprint
  {http://arxiv.org/abs/https://doi.org/10.1143/JPSJ.78.074707}
  {https://doi.org/10.1143/JPSJ.78.074707} \BibitemShut {NoStop}%
\bibitem [{\citenamefont {Jeong}\ and\ \citenamefont {Lee}(2009)}]{alphaTag2}%
  \BibitemOpen
  \bibfield  {author} {\bibinfo {author} {\bibfnamefont {J.-S.}\ \bibnamefont
  {Jeong}}\ and\ \bibinfo {author} {\bibfnamefont {H.-W.}\ \bibnamefont
  {Lee}},\ }\href {\doibase 10.1103/PhysRevB.80.075409} {\bibfield  {journal}
  {\bibinfo  {journal} {Phys. Rev. B}\ }\textbf {\bibinfo {volume} {80}},\
  \bibinfo {pages} {075409} (\bibinfo {year} {2009})}\BibitemShut {NoStop}%
\bibitem [{\citenamefont {Zhu}\ \emph {et~al.}(2011)\citenamefont {Zhu},
  \citenamefont {Cheng},\ and\ \citenamefont {Schwingenschl\"ogl}}]{TMDsoc1}%
  \BibitemOpen
  \bibfield  {author} {\bibinfo {author} {\bibfnamefont {Z.~Y.}\ \bibnamefont
  {Zhu}}, \bibinfo {author} {\bibfnamefont {Y.~C.}\ \bibnamefont {Cheng}}, \
  and\ \bibinfo {author} {\bibfnamefont {U.}~\bibnamefont
  {Schwingenschl\"ogl}},\ }\href {\doibase 10.1103/PhysRevB.84.153402}
  {\bibfield  {journal} {\bibinfo  {journal} {Phys. Rev. B}\ }\textbf {\bibinfo
  {volume} {84}},\ \bibinfo {pages} {153402} (\bibinfo {year}
  {2011})}\BibitemShut {NoStop}%
\bibitem [{\citenamefont {Xiao}\ \emph {et~al.}(2012)\citenamefont {Xiao},
  \citenamefont {Liu}, \citenamefont {Feng}, \citenamefont {Xu},\ and\
  \citenamefont {Yao}}]{TMDsoc2}%
  \BibitemOpen
  \bibfield  {author} {\bibinfo {author} {\bibfnamefont {D.}~\bibnamefont
  {Xiao}}, \bibinfo {author} {\bibfnamefont {G.-B.}\ \bibnamefont {Liu}},
  \bibinfo {author} {\bibfnamefont {W.}~\bibnamefont {Feng}}, \bibinfo {author}
  {\bibfnamefont {X.}~\bibnamefont {Xu}}, \ and\ \bibinfo {author}
  {\bibfnamefont {W.}~\bibnamefont {Yao}},\ }\href {\doibase
  10.1103/PhysRevLett.108.196802} {\bibfield  {journal} {\bibinfo  {journal}
  {Phys. Rev. Lett.}\ }\textbf {\bibinfo {volume} {108}},\ \bibinfo {pages}
  {196802} (\bibinfo {year} {2012})}\BibitemShut {NoStop}%
\bibitem [{\citenamefont {Lu}\ \emph {et~al.}(2015)\citenamefont {Lu},
  \citenamefont {Zheliuk}, \citenamefont {Leermakers}, \citenamefont {Yuan},
  \citenamefont {Zeitler}, \citenamefont {Law},\ and\ \citenamefont
  {Ye}}]{Mos2IsingSC}%
  \BibitemOpen
  \bibfield  {author} {\bibinfo {author} {\bibfnamefont {J.~M.}\ \bibnamefont
  {Lu}}, \bibinfo {author} {\bibfnamefont {O.}~\bibnamefont {Zheliuk}},
  \bibinfo {author} {\bibfnamefont {I.}~\bibnamefont {Leermakers}}, \bibinfo
  {author} {\bibfnamefont {N.~F.~Q.}\ \bibnamefont {Yuan}}, \bibinfo {author}
  {\bibfnamefont {U.}~\bibnamefont {Zeitler}}, \bibinfo {author} {\bibfnamefont
  {K.~T.}\ \bibnamefont {Law}}, \ and\ \bibinfo {author} {\bibfnamefont
  {J.~T.}\ \bibnamefont {Ye}},\ }\href {\doibase 10.1126/science.aab2277}
  {\bibfield  {journal} {\bibinfo  {journal} {Science}\ }\textbf {\bibinfo
  {volume} {350}},\ \bibinfo {pages} {1353} (\bibinfo {year} {2015})},\ \Eprint
  {http://arxiv.org/abs/https://science.sciencemag.org/content/350/6266/1353.full.pdf}
  {https://science.sciencemag.org/content/350/6266/1353.full.pdf} \BibitemShut
  {NoStop}%
\bibitem [{\citenamefont {Saito}\ \emph {et~al.}(2015)\citenamefont {Saito},
  \citenamefont {Nakamura}, \citenamefont {Bahramy}, \citenamefont {Kohama},
  \citenamefont {Ye}, \citenamefont {Kasahara}, \citenamefont {Nakagawa},
  \citenamefont {Onga}, \citenamefont {Tokunaga}, \citenamefont {Nojima},
  \citenamefont {Yanase},\ and\ \citenamefont {Iwasa}}]{LargeHcMos2}%
  \BibitemOpen
  \bibfield  {author} {\bibinfo {author} {\bibfnamefont {Y.}~\bibnamefont
  {Saito}}, \bibinfo {author} {\bibfnamefont {Y.}~\bibnamefont {Nakamura}},
  \bibinfo {author} {\bibfnamefont {M.~S.}\ \bibnamefont {Bahramy}}, \bibinfo
  {author} {\bibfnamefont {Y.}~\bibnamefont {Kohama}}, \bibinfo {author}
  {\bibfnamefont {J.}~\bibnamefont {Ye}}, \bibinfo {author} {\bibfnamefont
  {Y.}~\bibnamefont {Kasahara}}, \bibinfo {author} {\bibfnamefont
  {Y.}~\bibnamefont {Nakagawa}}, \bibinfo {author} {\bibfnamefont
  {M.}~\bibnamefont {Onga}}, \bibinfo {author} {\bibfnamefont {M.}~\bibnamefont
  {Tokunaga}}, \bibinfo {author} {\bibfnamefont {T.}~\bibnamefont {Nojima}},
  \bibinfo {author} {\bibfnamefont {Y.}~\bibnamefont {Yanase}}, \ and\ \bibinfo
  {author} {\bibfnamefont {Y.}~\bibnamefont {Iwasa}},\ }\href
  {https://doi.org/10.1038/nphys3580} {\bibfield  {journal} {\bibinfo
  {journal} {Nature Physics}\ }\textbf {\bibinfo {volume} {12}},\ \bibinfo
  {pages} {144 EP } (\bibinfo {year} {2015})}\BibitemShut {NoStop}%
\bibitem [{\citenamefont {Xi}\ \emph {et~al.}(2015)\citenamefont {Xi},
  \citenamefont {Wang}, \citenamefont {Zhao}, \citenamefont {Park},
  \citenamefont {Law}, \citenamefont {Berger}, \citenamefont {Forr{\'o}},
  \citenamefont {Shan},\ and\ \citenamefont {Mak}}]{ISINGtHINnBSE2}%
  \BibitemOpen
  \bibfield  {author} {\bibinfo {author} {\bibfnamefont {X.}~\bibnamefont
  {Xi}}, \bibinfo {author} {\bibfnamefont {Z.}~\bibnamefont {Wang}}, \bibinfo
  {author} {\bibfnamefont {W.}~\bibnamefont {Zhao}}, \bibinfo {author}
  {\bibfnamefont {J.-H.}\ \bibnamefont {Park}}, \bibinfo {author}
  {\bibfnamefont {K.~T.}\ \bibnamefont {Law}}, \bibinfo {author} {\bibfnamefont
  {H.}~\bibnamefont {Berger}}, \bibinfo {author} {\bibfnamefont
  {L.}~\bibnamefont {Forr{\'o}}}, \bibinfo {author} {\bibfnamefont
  {J.}~\bibnamefont {Shan}}, \ and\ \bibinfo {author} {\bibfnamefont {K.~F.}\
  \bibnamefont {Mak}},\ }\href {https://doi.org/10.1038/nphys3538} {\bibfield
  {journal} {\bibinfo  {journal} {Nature Physics}\ }\textbf {\bibinfo {volume}
  {12}},\ \bibinfo {pages} {139 EP } (\bibinfo {year} {2015})}\BibitemShut
  {NoStop}%
\bibitem [{\citenamefont {Dvir}\ \emph {et~al.}(2018)\citenamefont {Dvir},
  \citenamefont {Massee}, \citenamefont {Attias}, \citenamefont {Khodas},
  \citenamefont {Aprili}, \citenamefont {Quay},\ and\ \citenamefont
  {Steinberg}}]{Nbse2SpectroscopySteinberg}%
  \BibitemOpen
  \bibfield  {author} {\bibinfo {author} {\bibfnamefont {T.}~\bibnamefont
  {Dvir}}, \bibinfo {author} {\bibfnamefont {F.}~\bibnamefont {Massee}},
  \bibinfo {author} {\bibfnamefont {L.}~\bibnamefont {Attias}}, \bibinfo
  {author} {\bibfnamefont {M.}~\bibnamefont {Khodas}}, \bibinfo {author}
  {\bibfnamefont {M.}~\bibnamefont {Aprili}}, \bibinfo {author} {\bibfnamefont
  {C.~H.~L.}\ \bibnamefont {Quay}}, \ and\ \bibinfo {author} {\bibfnamefont
  {H.}~\bibnamefont {Steinberg}},\ }\href {\doibase 10.1038/s41467-018-03000-w}
  {\bibfield  {journal} {\bibinfo  {journal} {Nature Communications}\ }\textbf
  {\bibinfo {volume} {9}},\ \bibinfo {pages} {598} (\bibinfo {year}
  {2018})}\BibitemShut {NoStop}%
\bibitem [{\citenamefont {de~la Barrera}\ \emph {et~al.}(2018)\citenamefont
  {de~la Barrera}, \citenamefont {Sinko}, \citenamefont {Gopalan},
  \citenamefont {Sivadas}, \citenamefont {Seyler}, \citenamefont {Watanabe},
  \citenamefont {Taniguchi}, \citenamefont {Tsen}, \citenamefont {Xu},
  \citenamefont {Xiao},\ and\ \citenamefont {Hunt}}]{Nbse2IsingSCtuningHunt}%
  \BibitemOpen
  \bibfield  {author} {\bibinfo {author} {\bibfnamefont {S.~C.}\ \bibnamefont
  {de~la Barrera}}, \bibinfo {author} {\bibfnamefont {M.~R.}\ \bibnamefont
  {Sinko}}, \bibinfo {author} {\bibfnamefont {D.~P.}\ \bibnamefont {Gopalan}},
  \bibinfo {author} {\bibfnamefont {N.}~\bibnamefont {Sivadas}}, \bibinfo
  {author} {\bibfnamefont {K.~L.}\ \bibnamefont {Seyler}}, \bibinfo {author}
  {\bibfnamefont {K.}~\bibnamefont {Watanabe}}, \bibinfo {author}
  {\bibfnamefont {T.}~\bibnamefont {Taniguchi}}, \bibinfo {author}
  {\bibfnamefont {A.~W.}\ \bibnamefont {Tsen}}, \bibinfo {author}
  {\bibfnamefont {X.}~\bibnamefont {Xu}}, \bibinfo {author} {\bibfnamefont
  {D.}~\bibnamefont {Xiao}}, \ and\ \bibinfo {author} {\bibfnamefont {B.~M.}\
  \bibnamefont {Hunt}},\ }\href {\doibase 10.1038/s41467-018-03888-4}
  {\bibfield  {journal} {\bibinfo  {journal} {Nature Communications}\ }\textbf
  {\bibinfo {volume} {9}},\ \bibinfo {pages} {1427} (\bibinfo {year}
  {2018})}\BibitemShut {NoStop}%
\bibitem [{\citenamefont {Sohn}\ \emph {et~al.}(2018)\citenamefont {Sohn},
  \citenamefont {Xi}, \citenamefont {He}, \citenamefont {Jiang}, \citenamefont
  {Wang}, \citenamefont {Kang}, \citenamefont {Park}, \citenamefont {Berger},
  \citenamefont {Forr{\'o}}, \citenamefont {Law}, \citenamefont {Shan},\ and\
  \citenamefont {Mak}}]{NbSe2Mak}%
  \BibitemOpen
  \bibfield  {author} {\bibinfo {author} {\bibfnamefont {E.}~\bibnamefont
  {Sohn}}, \bibinfo {author} {\bibfnamefont {X.}~\bibnamefont {Xi}}, \bibinfo
  {author} {\bibfnamefont {W.-Y.}\ \bibnamefont {He}}, \bibinfo {author}
  {\bibfnamefont {S.}~\bibnamefont {Jiang}}, \bibinfo {author} {\bibfnamefont
  {Z.}~\bibnamefont {Wang}}, \bibinfo {author} {\bibfnamefont {K.}~\bibnamefont
  {Kang}}, \bibinfo {author} {\bibfnamefont {J.-H.}\ \bibnamefont {Park}},
  \bibinfo {author} {\bibfnamefont {H.}~\bibnamefont {Berger}}, \bibinfo
  {author} {\bibfnamefont {L.}~\bibnamefont {Forr{\'o}}}, \bibinfo {author}
  {\bibfnamefont {K.~T.}\ \bibnamefont {Law}}, \bibinfo {author} {\bibfnamefont
  {J.}~\bibnamefont {Shan}}, \ and\ \bibinfo {author} {\bibfnamefont {K.~F.}\
  \bibnamefont {Mak}},\ }\href {\doibase 10.1038/s41563-018-0061-1} {\bibfield
  {journal} {\bibinfo  {journal} {Nature Materials}\ }\textbf {\bibinfo
  {volume} {17}},\ \bibinfo {pages} {504} (\bibinfo {year} {2018})}\BibitemShut
  {NoStop}%
\bibitem [{\citenamefont {Zhou}\ \emph {et~al.}(2016)\citenamefont {Zhou},
  \citenamefont {Yuan}, \citenamefont {Jiang},\ and\ \citenamefont
  {Law}}]{TMDhalfmetal}%
  \BibitemOpen
  \bibfield  {author} {\bibinfo {author} {\bibfnamefont {B.~T.}\ \bibnamefont
  {Zhou}}, \bibinfo {author} {\bibfnamefont {N.~F.~Q.}\ \bibnamefont {Yuan}},
  \bibinfo {author} {\bibfnamefont {H.-L.}\ \bibnamefont {Jiang}}, \ and\
  \bibinfo {author} {\bibfnamefont {K.~T.}\ \bibnamefont {Law}},\ }\href
  {\doibase 10.1103/PhysRevB.93.180501} {\bibfield  {journal} {\bibinfo
  {journal} {Phys. Rev. B}\ }\textbf {\bibinfo {volume} {93}},\ \bibinfo
  {pages} {180501} (\bibinfo {year} {2016})}\BibitemShut {NoStop}%
\bibitem [{\citenamefont {Kim}\ \emph {et~al.}(2017)\citenamefont {Kim},
  \citenamefont {Park}, \citenamefont {Lee}, \citenamefont {Lee}, \citenamefont
  {Park}, \citenamefont {Lee}, \citenamefont {Lee},\ and\ \citenamefont
  {Lee}}]{Nbse2GrapheneJunction2}%
  \BibitemOpen
  \bibfield  {author} {\bibinfo {author} {\bibfnamefont {M.}~\bibnamefont
  {Kim}}, \bibinfo {author} {\bibfnamefont {G.-H.}\ \bibnamefont {Park}},
  \bibinfo {author} {\bibfnamefont {J.}~\bibnamefont {Lee}}, \bibinfo {author}
  {\bibfnamefont {J.~H.}\ \bibnamefont {Lee}}, \bibinfo {author} {\bibfnamefont
  {J.}~\bibnamefont {Park}}, \bibinfo {author} {\bibfnamefont {H.}~\bibnamefont
  {Lee}}, \bibinfo {author} {\bibfnamefont {G.-H.}\ \bibnamefont {Lee}}, \ and\
  \bibinfo {author} {\bibfnamefont {H.-J.}\ \bibnamefont {Lee}},\ }\href
  {\doibase 10.1021/acs.nanolett.7b02707} {\bibfield  {journal} {\bibinfo
  {journal} {Nano Letters}\ }\textbf {\bibinfo {volume} {17}},\ \bibinfo
  {pages} {6125} (\bibinfo {year} {2017})},\ \bibinfo {note} {pMID: 28952735},\
  \Eprint {http://arxiv.org/abs/https://doi.org/10.1021/acs.nanolett.7b02707}
  {https://doi.org/10.1021/acs.nanolett.7b02707} \BibitemShut {NoStop}%
\bibitem [{\citenamefont {Han}\ \emph {et~al.}(2018)\citenamefont {Han},
  \citenamefont {Shen}, \citenamefont {Yuan}, \citenamefont {Lin},
  \citenamefont {Wu}, \citenamefont {Wu}, \citenamefont {Xu}, \citenamefont
  {An}, \citenamefont {Long}, \citenamefont {Wang}, \citenamefont {Lortz},\
  and\ \citenamefont {Wang}}]{NbSe2garaphene3}%
  \BibitemOpen
  \bibfield  {author} {\bibinfo {author} {\bibfnamefont {T.}~\bibnamefont
  {Han}}, \bibinfo {author} {\bibfnamefont {J.}~\bibnamefont {Shen}}, \bibinfo
  {author} {\bibfnamefont {N.~F.~Q.}\ \bibnamefont {Yuan}}, \bibinfo {author}
  {\bibfnamefont {J.}~\bibnamefont {Lin}}, \bibinfo {author} {\bibfnamefont
  {Z.}~\bibnamefont {Wu}}, \bibinfo {author} {\bibfnamefont {Y.}~\bibnamefont
  {Wu}}, \bibinfo {author} {\bibfnamefont {S.}~\bibnamefont {Xu}}, \bibinfo
  {author} {\bibfnamefont {L.}~\bibnamefont {An}}, \bibinfo {author}
  {\bibfnamefont {G.}~\bibnamefont {Long}}, \bibinfo {author} {\bibfnamefont
  {Y.}~\bibnamefont {Wang}}, \bibinfo {author} {\bibfnamefont {R.}~\bibnamefont
  {Lortz}}, \ and\ \bibinfo {author} {\bibfnamefont {N.}~\bibnamefont {Wang}},\
  }\href {\doibase 10.1103/PhysRevB.97.060505} {\bibfield  {journal} {\bibinfo
  {journal} {Phys. Rev. B}\ }\textbf {\bibinfo {volume} {97}},\ \bibinfo
  {pages} {060505} (\bibinfo {year} {2018})}\BibitemShut {NoStop}%
\bibitem [{\citenamefont {Yarimizu}\ \emph {et~al.}(2018)\citenamefont
  {Yarimizu}, \citenamefont {Tomori}, \citenamefont {Watanabe}, \citenamefont
  {Taniguchi},\ and\ \citenamefont {Kanda}}]{Nbse2GrapheneJunction}%
  \BibitemOpen
  \bibfield  {author} {\bibinfo {author} {\bibfnamefont {K.}~\bibnamefont
  {Yarimizu}}, \bibinfo {author} {\bibfnamefont {H.}~\bibnamefont {Tomori}},
  \bibinfo {author} {\bibfnamefont {K.}~\bibnamefont {Watanabe}}, \bibinfo
  {author} {\bibfnamefont {T.}~\bibnamefont {Taniguchi}}, \ and\ \bibinfo
  {author} {\bibfnamefont {A.}~\bibnamefont {Kanda}},\ }\href {\doibase
  10.1088/1742-6596/969/1/012147} {\bibfield  {journal} {\bibinfo  {journal}
  {Journal of Physics: Conference Series}\ }\textbf {\bibinfo {volume} {969}},\
  \bibinfo {pages} {012147} (\bibinfo {year} {2018})}\BibitemShut {NoStop}%
\bibitem [{\citenamefont {Ghosh}\ \emph {et~al.}(2010)\citenamefont {Ghosh},
  \citenamefont {Sau}, \citenamefont {Tewari},\ and\ \citenamefont
  {Das~Sarma}}]{PfaffianInvariant}%
  \BibitemOpen
  \bibfield  {author} {\bibinfo {author} {\bibfnamefont {P.}~\bibnamefont
  {Ghosh}}, \bibinfo {author} {\bibfnamefont {J.~D.}\ \bibnamefont {Sau}},
  \bibinfo {author} {\bibfnamefont {S.}~\bibnamefont {Tewari}}, \ and\ \bibinfo
  {author} {\bibfnamefont {S.}~\bibnamefont {Das~Sarma}},\ }\href {\doibase
  10.1103/PhysRevB.82.184525} {\bibfield  {journal} {\bibinfo  {journal} {Phys.
  Rev. B}\ }\textbf {\bibinfo {volume} {82}},\ \bibinfo {pages} {184525}
  (\bibinfo {year} {2010})}\BibitemShut {NoStop}%
\bibitem [{\citenamefont {Lutchyn}\ \emph {et~al.}(2011)\citenamefont
  {Lutchyn}, \citenamefont {Stanescu},\ and\ \citenamefont
  {Das~Sarma}}]{PfaffianMultiband}%
  \BibitemOpen
  \bibfield  {author} {\bibinfo {author} {\bibfnamefont {R.~M.}\ \bibnamefont
  {Lutchyn}}, \bibinfo {author} {\bibfnamefont {T.~D.}\ \bibnamefont
  {Stanescu}}, \ and\ \bibinfo {author} {\bibfnamefont {S.}~\bibnamefont
  {Das~Sarma}},\ }\href {\doibase 10.1103/PhysRevLett.106.127001} {\bibfield
  {journal} {\bibinfo  {journal} {Phys. Rev. Lett.}\ }\textbf {\bibinfo
  {volume} {106}},\ \bibinfo {pages} {127001} (\bibinfo {year}
  {2011})}\BibitemShut {NoStop}%
\bibitem [{\citenamefont {Peng}\ \emph {et~al.}(2015)\citenamefont {Peng},
  \citenamefont {Pientka}, \citenamefont {Glazman},\ and\ \citenamefont {{von
  Oppen}}}]{peng_strong_2015}%
  \BibitemOpen
  \bibfield  {author} {\bibinfo {author} {\bibfnamefont {Y.}~\bibnamefont
  {Peng}}, \bibinfo {author} {\bibfnamefont {F.}~\bibnamefont {Pientka}},
  \bibinfo {author} {\bibfnamefont {L.~I.}\ \bibnamefont {Glazman}}, \ and\
  \bibinfo {author} {\bibfnamefont {F.}~\bibnamefont {{von Oppen}}},\ }\href
  {\doibase 10.1103/PhysRevLett.114.106801} {\bibfield  {journal} {\bibinfo
  {journal} {Physical Review Letters}\ }\textbf {\bibinfo {volume} {114}},\
  \bibinfo {pages} {106801} (\bibinfo {year} {2015})}\BibitemShut {NoStop}%
\bibitem [{\citenamefont {Tom\'anek}\ and\ \citenamefont
  {Louie}(1988)}]{tomanek_first-principles_1988}%
  \BibitemOpen
  \bibfield  {author} {\bibinfo {author} {\bibfnamefont {D.}~\bibnamefont
  {Tom\'anek}}\ and\ \bibinfo {author} {\bibfnamefont {S.~G.}\ \bibnamefont
  {Louie}},\ }\href {\doibase 10.1103/PhysRevB.37.8327} {\bibfield  {journal}
  {\bibinfo  {journal} {Physical Review B}\ }\textbf {\bibinfo {volume} {37}},\
  \bibinfo {pages} {8327} (\bibinfo {year} {1988})}\BibitemShut {NoStop}%
\bibitem [{\citenamefont {Haim}\ \emph {et~al.}(2016)\citenamefont {Haim},
  \citenamefont {Berg}, \citenamefont {Flensberg},\ and\ \citenamefont
  {Oreg}}]{NoGoMAJORANA}%
  \BibitemOpen
  \bibfield  {author} {\bibinfo {author} {\bibfnamefont {A.}~\bibnamefont
  {Haim}}, \bibinfo {author} {\bibfnamefont {E.}~\bibnamefont {Berg}}, \bibinfo
  {author} {\bibfnamefont {K.}~\bibnamefont {Flensberg}}, \ and\ \bibinfo
  {author} {\bibfnamefont {Y.}~\bibnamefont {Oreg}},\ }\href {\doibase
  10.1103/PhysRevB.94.161110} {\bibfield  {journal} {\bibinfo  {journal} {Phys.
  Rev. B}\ }\textbf {\bibinfo {volume} {94}},\ \bibinfo {pages} {161110}
  (\bibinfo {year} {2016})}\BibitemShut {NoStop}%
\bibitem [{\citenamefont {Haim}\ \emph {et~al.}(2014)\citenamefont {Haim},
  \citenamefont {Keselman}, \citenamefont {Berg},\ and\ \citenamefont
  {Oreg}}]{TritopsHFpaper}%
  \BibitemOpen
  \bibfield  {author} {\bibinfo {author} {\bibfnamefont {A.}~\bibnamefont
  {Haim}}, \bibinfo {author} {\bibfnamefont {A.}~\bibnamefont {Keselman}},
  \bibinfo {author} {\bibfnamefont {E.}~\bibnamefont {Berg}}, \ and\ \bibinfo
  {author} {\bibfnamefont {Y.}~\bibnamefont {Oreg}},\ }\href {\doibase
  10.1103/PhysRevB.89.220504} {\bibfield  {journal} {\bibinfo  {journal} {Phys.
  Rev. B}\ }\textbf {\bibinfo {volume} {89}},\ \bibinfo {pages} {220504}
  (\bibinfo {year} {2014})}\BibitemShut {NoStop}%
\bibitem [{\citenamefont {Kane}\ \emph {et~al.}(1997)\citenamefont {Kane},
  \citenamefont {Balents},\ and\ \citenamefont {Fisher}}]{CNTinteractionVr}%
  \BibitemOpen
  \bibfield  {author} {\bibinfo {author} {\bibfnamefont {C.}~\bibnamefont
  {Kane}}, \bibinfo {author} {\bibfnamefont {L.}~\bibnamefont {Balents}}, \
  and\ \bibinfo {author} {\bibfnamefont {M.~P.~A.}\ \bibnamefont {Fisher}},\
  }\href {\doibase 10.1103/PhysRevLett.79.5086} {\bibfield  {journal} {\bibinfo
   {journal} {Phys. Rev. Lett.}\ }\textbf {\bibinfo {volume} {79}},\ \bibinfo
  {pages} {5086} (\bibinfo {year} {1997})}\BibitemShut {NoStop}%
\bibitem [{\citenamefont {Cao}\ \emph {et~al.}(2018)\citenamefont {Cao},
  \citenamefont {Fatemi}, \citenamefont {Fang}, \citenamefont {Watanabe},
  \citenamefont {Taniguchi}, \citenamefont {Kaxiras},\ and\ \citenamefont
  {Jarillo-Herrero}}]{MITtblg}%
  \BibitemOpen
  \bibfield  {author} {\bibinfo {author} {\bibfnamefont {Y.}~\bibnamefont
  {Cao}}, \bibinfo {author} {\bibfnamefont {V.}~\bibnamefont {Fatemi}},
  \bibinfo {author} {\bibfnamefont {S.}~\bibnamefont {Fang}}, \bibinfo {author}
  {\bibfnamefont {K.}~\bibnamefont {Watanabe}}, \bibinfo {author}
  {\bibfnamefont {T.}~\bibnamefont {Taniguchi}}, \bibinfo {author}
  {\bibfnamefont {E.}~\bibnamefont {Kaxiras}}, \ and\ \bibinfo {author}
  {\bibfnamefont {P.}~\bibnamefont {Jarillo-Herrero}},\ }\href {\doibase
  10.1038/nature26160} {\bibfield  {journal} {\bibinfo  {journal} {Nature}\
  }\textbf {\bibinfo {volume} {556}},\ \bibinfo {pages} {43} (\bibinfo {year}
  {2018})}\BibitemShut {NoStop}%
\bibitem [{\citenamefont {You}\ and\ \citenamefont
  {Vishwanath}(2019)}]{TBLGtriplet}%
  \BibitemOpen
  \bibfield  {author} {\bibinfo {author} {\bibfnamefont {Y.-Z.}\ \bibnamefont
  {You}}\ and\ \bibinfo {author} {\bibfnamefont {A.}~\bibnamefont
  {Vishwanath}},\ }\href {\doibase 10.1038/s41535-019-0153-4} {\bibfield
  {journal} {\bibinfo  {journal} {npj Quantum Materials}\ }\textbf {\bibinfo
  {volume} {4}},\ \bibinfo {pages} {16} (\bibinfo {year} {2019})}\BibitemShut
  {NoStop}%
\bibitem [{\citenamefont {Efroni}\ \emph {et~al.}(2017)\citenamefont {Efroni},
  \citenamefont {Ilani},\ and\ \citenamefont {Berg}}]{efroni_topological_2017}%
  \BibitemOpen
  \bibfield  {author} {\bibinfo {author} {\bibfnamefont {Y.}~\bibnamefont
  {Efroni}}, \bibinfo {author} {\bibfnamefont {S.}~\bibnamefont {Ilani}}, \
  and\ \bibinfo {author} {\bibfnamefont {E.}~\bibnamefont {Berg}},\ }\href
  {\doibase 10.1103/PhysRevLett.119.147704} {\bibfield  {journal} {\bibinfo
  {journal} {Physical Review Letters}\ }\textbf {\bibinfo {volume} {119}},\
  \bibinfo {pages} {147704} (\bibinfo {year} {2017})}\BibitemShut {NoStop}%
\bibitem [{\citenamefont {Giamarchi}\ and\ \citenamefont
  {Press}(2004)}]{giamarchi2004quantum}%
  \BibitemOpen
  \bibfield  {author} {\bibinfo {author} {\bibfnamefont {T.}~\bibnamefont
  {Giamarchi}}\ and\ \bibinfo {author} {\bibfnamefont {O.~U.}\ \bibnamefont
  {Press}},\ }\href {https://books.google.co.il/books?id=1MwTDAAAQBAJ} {\emph
  {\bibinfo {title} {Quantum Physics in One Dimension}}},\ International Series
  of Monogr\ (\bibinfo  {publisher} {Clarendon Press},\ \bibinfo {year}
  {2004})\BibitemShut {NoStop}%
\bibitem [{Note1()}]{Note1}%
  \BibitemOpen
  \bibinfo {note} {$g$ is the usual backscattering $g_{1\perp }$ term in the
  standard g-ology language.}\BibitemShut {Stop}%
\bibitem [{\citenamefont {Cardy}(1996)}]{cardy_1996}%
  \BibitemOpen
  \bibfield  {author} {\bibinfo {author} {\bibfnamefont {J.}~\bibnamefont
  {Cardy}},\ }\href {\doibase 10.1017/CBO9781316036440} {\emph {\bibinfo
  {title} {Scaling and Renormalization in Statistical Physics}}},\ Cambridge
  Lecture Notes in Physics\ (\bibinfo  {publisher} {Cambridge University
  Press},\ \bibinfo {year} {1996})\BibitemShut {NoStop}%
\end{thebibliography}%

\end{document}